\documentclass[aps,prb,reprint]{revtex4-2}
\usepackage{graphicx} % Include figure files
\usepackage{amsmath} % Useful mathematical tools
\usepackage{bm} % bold mathi
\usepackage{natbib}
\usepackage{natmove}
\usepackage[colorlinks=true, linkcolor=blue, citecolor=blue, urlcolor=blue, linktoc=page, bookmarks=false, pdfstartview={FitH}, pdfborder={0 0 0.0 [3 3]}]{hyperref} % HyperLink
\usepackage{color} % For colored text
\usepackage{dcolumn} % Align table columns on decimal point
\newcolumntype{.}{D{.}{.}{1.3}}
\newcolumntype{-}{D{.}{.}{4.0}}
\newcolumntype{:}{D{:}{:}{12.30}}
\newcolumntype{*}{D{:}{:}{1.7}}
\usepackage{makecell}
\usepackage{multirow}
\usepackage{cleveref}
\usepackage{easyReview} % Review mode
\crefname{figure}{Fig.}{Figs}
\crefname{table}{Table}{Tables}
\crefname{section}{Sec.}{Secs.}
\crefname{equation}{Eq.}{Eqs.}
\renewcommand{\today}{\number\day \space \ifcase \month \or January\or February\or March\or April\or May\or June\or July\or August\or September\or October\or November\or December\fi \space \number\year} % Date
\def\m1r{\multicolumn{1}{r}}
\begin{document}
% ========== TITLE ==========
\title{Altermagnetism in orthorhombic $Pnma$ structure through group theory and DFT calculations}
% ====================
% ========== AUTHORS AND AFFILIATIONS ==========
\author{Suman \surname{Rooj}\add{$^{\dagger}$}}
%\thanks{These two authors contributed equally}
\email[Email: ]{suman18@iiserb.ac.in}
%-------------------------------------------------
\author{Sugandha \surname{Saxena}}
\thanks{These two authors contributed equally}
%--------------------------------------------------------
\author{Nirmal \surname{Ganguli}}
\email[Email: ]{NGanguli@iiserb.ac.in}
\affiliation{Department of Physics, \href{https://ror.org/02rb21j89}{Indian Institute of Science Education and Research Bhopal}, Bhauri, Bhopal 462066, India}
% ====================
\date{\today}
% ========== ABSTRACT ==========
\begin{abstract}
Antiferromagnetism, initially considered interesting but useless, recently emerged as one of the most promising magnetic phases for technology. Recently, a low symmetry antiferromagnetic phase, known as altermagnetic phase, have been discovered, where no time reversal ($\mathcal{T}$) symmetry is observed in spite of a vanishing net magnetization, leading to non-degenerate bands from the opposite magnetic sublattices. In this work, we consider two representatives of orthorhombic $Pnma$ space group, namely, BiFeO$_3$ and CaMnO$_3$, and find altermagnetic lowest energy phase in both from our density functional theory calculations. We find a substantial spin-splitting in both systems along a high-symmetry path in the Brillouin zone without considering the spin-orbit interaction (SOI). Detailed features of the band dispersion obtained from our calculation confirm the lifting of sublattice spin degeneracy only in the $k_x = 0$ plane while preserving the spin degeneracy in the other planes of the Brillouin zone. We provide a comprehensive symmetry analysis based on the magnetic space group (MSG) to explain our DFT findings and an insightful symmetry-allowed model Hamiltonian, which qualitatively agrees with our results. Additionally, we extend our symmetry analysis to encompass two other potential MSGs within the $Pnma$ space group that may host the spin-splitting phenomenon without considering SOI and the likely form of their Hamiltonian. Our calculations considering SOI reveal weak ferromagnetism in both systems. The detailed studies presented here pave the way for a deeper understanding of the spin-splitting phenomena within the $Pnma$ space group, offering insights into the intricate interplay between symmetry and electronic as well as magnetic properties.
\end{abstract}
% ====================
%\pacs{} % Physics and Astronomy Classification Scheme
\maketitle

% ========== INTRODUCTION ==========
\section{\label{sec:intro}Introduction}
Magnetism in quantum materials and heterostructures exhibits fascinating scientific phenomena besides presenting exciting possibilities for new technology. The interplay between three fundamental ingredients, magnetism, spin-orbit interaction, and crystal symmetry, may lead to many exotic features in a quantum material, including topologically nontrivial properties, Rashba-Dresselhaus interaction, spin-Hall effect \cite{KeimerNP17, Manchon2015}. Among these, antiferromagnetic spintronics emerges as a promising frontier, wherein materials with significant spin-orbit interaction enable THz-speed computational processing and nonvolatile memory storage through spin-texture manipulation via spin-orbit torque \cite{RevModPhys.90.015005}. However, this paradigm undergoes a significant shift following the theoretical prediction and experimental verification of an interesting collinear antiferromagnetic phase, known as altermagnets, where the bands from the opposite magnetic sublattices are non-degenerate \cite{RoojAPR23, Libor2020, PhysRevX.12.031042, PhysRevX.12.040501, PhysRevMaterials.5.014409, PhysRevB.102.014422, Hayami2020}. Altermagnetic materials present fascinating attributes where time-reversal symmetry-breaking responses and spin-polarization emerge independent of spin-orbit interaction, accompanied by the collinear antiparallel magnetic order, leading to a vanishing net magnetization \cite{PhysRevX.12.031042, PhysRevX.12.040501}. RuO$_2$, an earlier proposed material for altermagnetism, showcases spin-polarization, resulting in a spin-splitter torque, which holds promise for manipulating spin-textures \cite{Hernandez2021, HBai2022}, although recent experiments reveal no magnetic order in RuO$_2$ \cite{KesslerNPJS24, LiuPRL24}. Besides the spin-splitter torque, some of the altermagnetic materials have been theoretically predicted and experimentally validated to host various other interesting phenomena. For instance, the anomalous Hall effect (AHE) has been experimentally demonstrated in some altermagnets such as MnTe, Mn$_5$Si$_3$, and RuO$_2$ \cite{MnTeAHE2023, Feng2022, kluczyk2023, leiviskä2024}. Additionally, giant tunneling magnetoresistance has been theoretically predicted in the RuO$_2|\text{TiO}_2|\text{RuO}_2$ heterostructure \cite{Jiang2023}. Furthermore, theoretical predictions of chiral magnons, crystal Nernst, and crystal thermal Hall effect have been reported in RuO$_2$ \cite{LiborMagnon2023, Libor2024}. Experimental observations have been made on the giant spin polarization in the non-relativistic limit in MnTe \cite{PhysRevB.109.115102}. With substantial spin-orbit interaction, altermagnets often exhibit a coexisting weak ferromagnetism and anomalous Hall effect \cite{Milivojevic2DM24, FakhredinePRB23, SinghARXIV24}. Antiferromagnetic hourglass electrons and nodal lines may also be found in such materials in the presence of nonsymmorphic symmetries \cite{FakhredinePRB23}. These collective insights illustrate the emerging landscape of altermagnetism, perpetuating the ongoing pursuit of suitable altermagnetic materials. 

In this work, based on the crystal symmetry, we identify two centrosymmetric materials: bulk orthorhombic BiFeO$_3$ (BFO) and CaMnO$_3$ (CMO), both belonging to nonsymmorphic $Pnma$ space group, as potential altermagnets. Employing first-principles density functional theory (DFT) calculations, we investigate the electronic structure of BFO and CMO in their orthorhombic structure. After finding the ground state magnetic arrangement for both systems, we critically examine the spin-splitting behavior with and without spin-orbit interaction within DFT calculations. In this study, we primarily focus on spin-splitting behavior without spin-orbit interaction. Additionally, we explain our findings using magnetic space group (MSG) symmetry analysis and an insightful analytical model. Our prediction of the spin-splitting effects without spin-orbit interaction conclusively designates these two compounds as altermagnets. We further anticipate the possible spin-splitting behavior for two other potential MSGs within the $Pnma$ space group and the probable form of model Hamiltonians. Our investigation based on DFT calculations and symmetry analysis including spin-orbit interaction reveals the presence of a weak ferromagnetic component in these compounds. Therefore, we explore the possibility of anomalous Hall effect in the system. The remainder of the article is structured as follows: The crystal structure and calculation methodologies are described in \cref{sec:method}. In \cref{sec:Results}, we rigorously discuss the results encompassing the electronic structure and symmetry analysis of our investigation. Finally, we summarize our work in \cref{sec:discussion}.

% ========== METHOD ==========
\section{\label{sec:method}Crystal Structure and Methodology}
\begin{table}
\caption{\label{tab:LatticeConstant}The lattice constants of BiFeO$_3$ and CaMnO$_3$, as obtained from experiments are tabulated here \cite{bfo_lattice,cmo_lattice}.}
\begin{ruledtabular}
    \begin{tabular}{l...}
     Crystal structure & \multicolumn{1}{c}{$a$~(\AA)} & \multicolumn{1}{c}{$b$~(\AA)} & \multicolumn{1}{c}{$c$~(\AA)} \\
    \hline
    BiFeO$_3$ & 5.61 & 7.97 & 5.65\\
    CaMnO$_3$ & 5.33 & 7.50 & 5.31\\
    \end{tabular}
\end{ruledtabular}
\end{table}

\begin{figure*}
\includegraphics[scale=0.72]{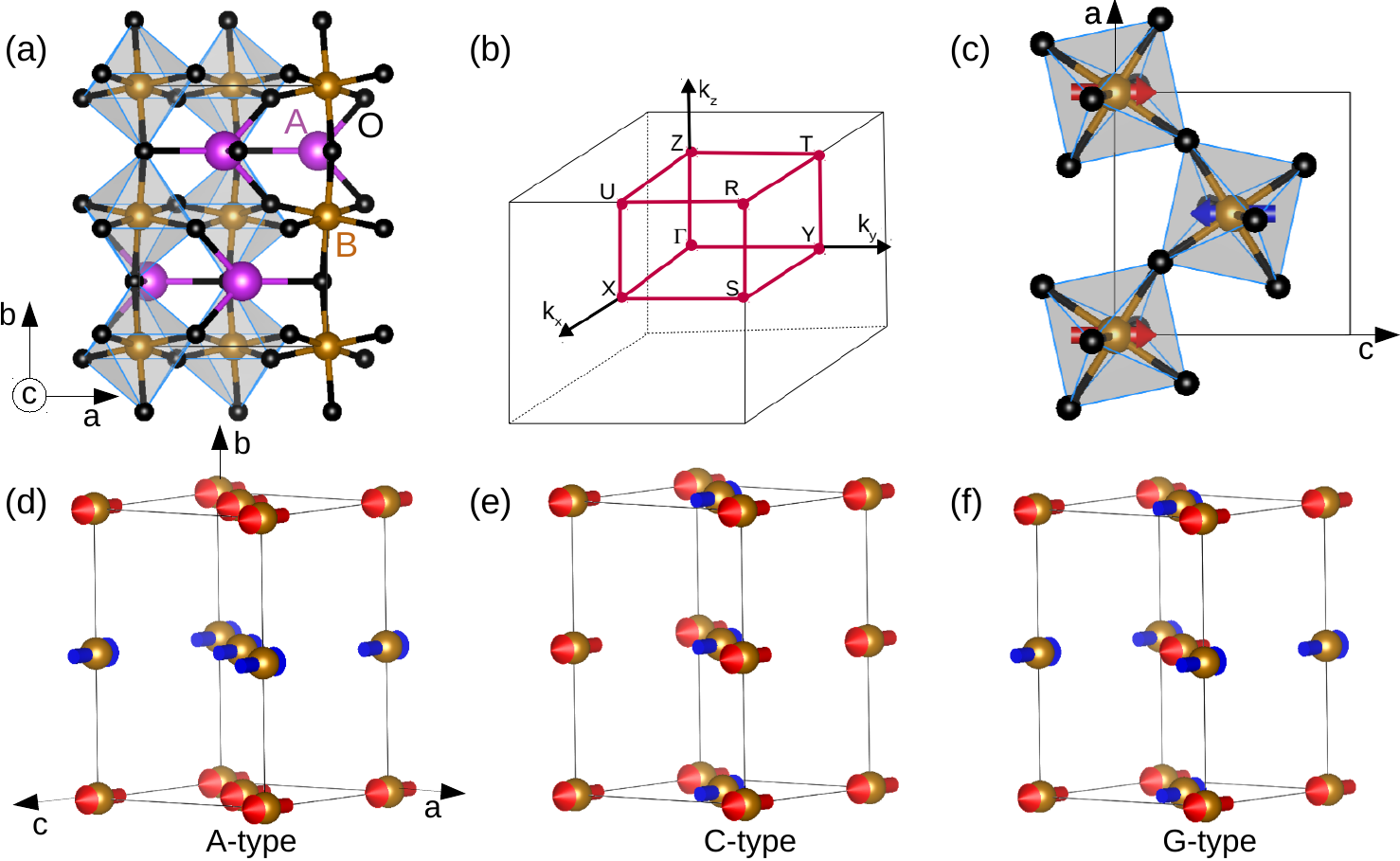}
	\caption{\label{Fig:structure} Panel (a) and (b) display an orthorhombic unit cell of ABO$_3$ Perovskite oxide structure, highlighting the corner-sharing BO$_6$ octahedra and the Brillouin zone, respectively, while (c) depicts a view on the $ac$ plane, illustrating the broken $U\tau (\frac{1}{2}0\frac{1}{2})$ symmetry. The A, C, and G-type antiferromagnetic configurations are illustrated in (d), (e), and (f), respectively. Crystal structure and magnetic configuration illustrations were prepared using {\scshape vesta} software \cite{VESTA}.} 
\end{figure*}

Bulk BiFeO$_3$ (BFO) and CaMnO$_3$ (CMO) crystallize in a centrosymmetric orthorhombic structure with $Pnma$ space group. The Perovskite oxides BFO and CMO may be described by a general formula ABO$_\text{3}$, with A = Bi/Ca, B = Fe/Mn. The A, B, and O sites occupy the Wyckoff positions $4c$, $4a$, and ($8d, 4c$), respectively \cite{Bilbao2006, cmo_wyckoff}. \Cref{Fig:structure}(a) depicts a unit cell of ABO$_\text{3}$, illustrating corner sharing BO$_6$ octahedra, while A atoms occupy the interstitial sites. The experimental lattice parameters for both compounds are tabulated in \cref{tab:LatticeConstant}. The total energy, electronic structure, magnetic properties, and spin-orbit interaction calculations are carried out within density functional theory (DFT) as implemented in the {\scshape vasp} code \cite{vasp1,vasp2}. The projector augmented wave (PAW) method is employed for the potential description alongside a plane wave basis set with a 500~eV energy cutoff for expanding the wavefunctions \cite{paw}. The exchange-correlation functional is treated within the generalized gradient approximation (GGA) \cite{pbe} with Hubbard-$U$ \cite{DudarevPRB98} correction of $U_\text{eff}= U-J=4~\text{and}~3$~eV for Fe-$3d$ and Mn-$3d$ states respectively to take care of the strong Coulomb correlation in BFO and CMO \cite{BFO_ueff, CMO_ueff}. The integration over the Brillouin zone is performed using a $5 \times 3 \times 5$ $\Gamma$-centered $k$-point mesh for the unit cell within the corrected tetrahedron method \cite{BlochTetrahedron94}. The electronic convergence threshold of $10^{-6}$~eV is employed for all calculations. Atomic positions and lattice vectors are optimized by minimizing the Hellman-Feynman force on each atom up to a threshold value of $10^{-2}$~eV/\AA. The anomalous Hall conductivity tensor is calculated using the Wannier90 code \cite{wannier90}.

% ========== RESULTS AND DISCUSSIONS ==========

\section{\label{sec:Results}Result and Discussion}
\subsection{\label{subsec:ElecandMagproperties}Electronic structure and magnetic properties}
%-----------------------Unit cell-----------------------------------
\begin{table}
\caption{\label{tab:Magnetic Configuration}Different magnetic configurations and their relative energies in the meV unit are listed here for the unit cell of both systems, BFO and CMO.}
\begin{ruledtabular}
    \begin{tabular}{l..}
     Magnetic Configuration & \multicolumn{1}{c}{BFO} & \multicolumn{1}{c}{CMO} \\
    \hline
      FM  & 825.97 & 184.49\\
      A - AFM & 515.04 &  72.92\\
      C - AFM & 233.63 &  31.34\\
      G - AFM & 0 & 0\\
    \end{tabular}
\end{ruledtabular}
\end{table}

\begin{figure}
\includegraphics[scale=0.44]{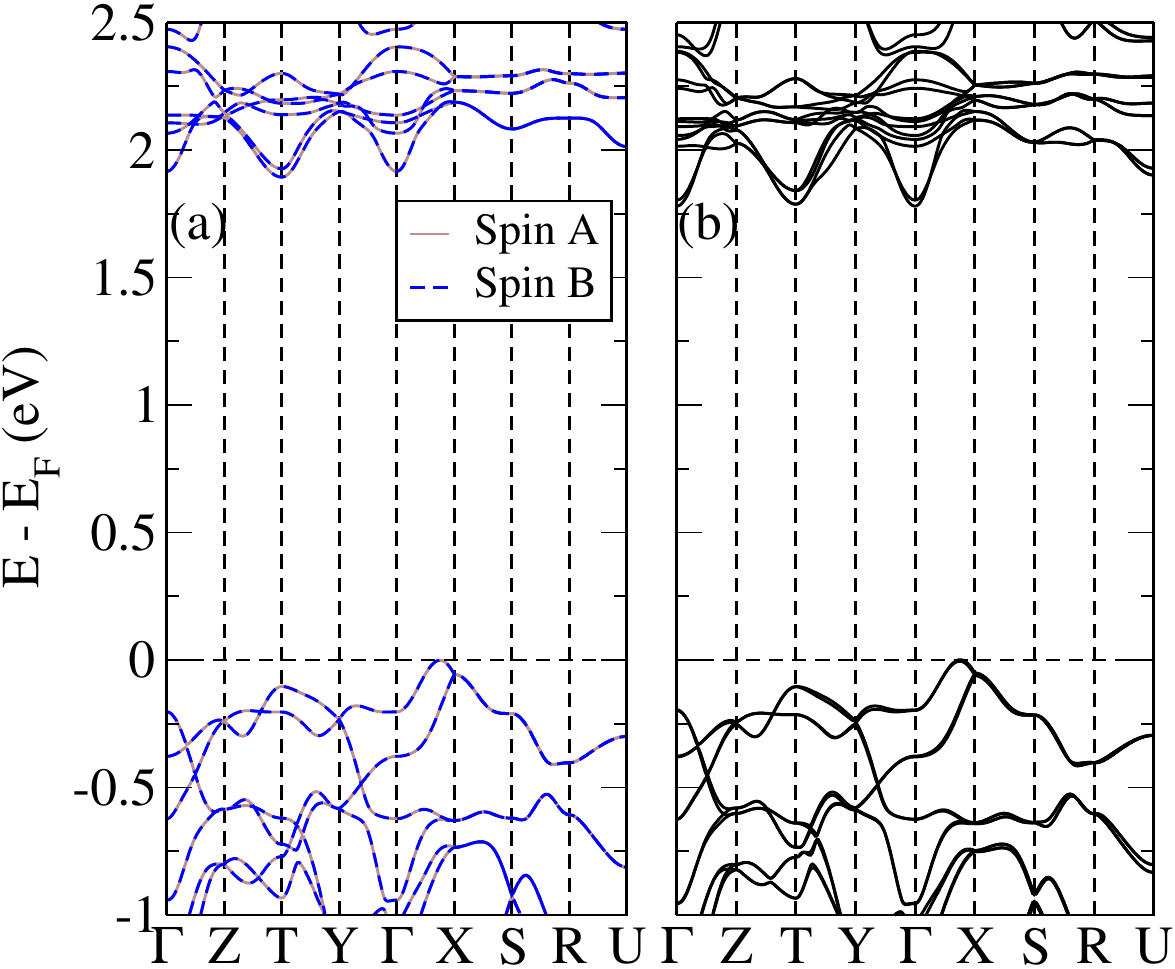}
	\caption{\label{Fig:BFOfullpath} Spin-polarized band dispersion for BFO from both spins marked as spin A and spin B is depicted in (a), while (b) represents the band structure considering spin-orbit interaction.}
\end{figure}

\begin{figure}
\includegraphics[scale=0.44]{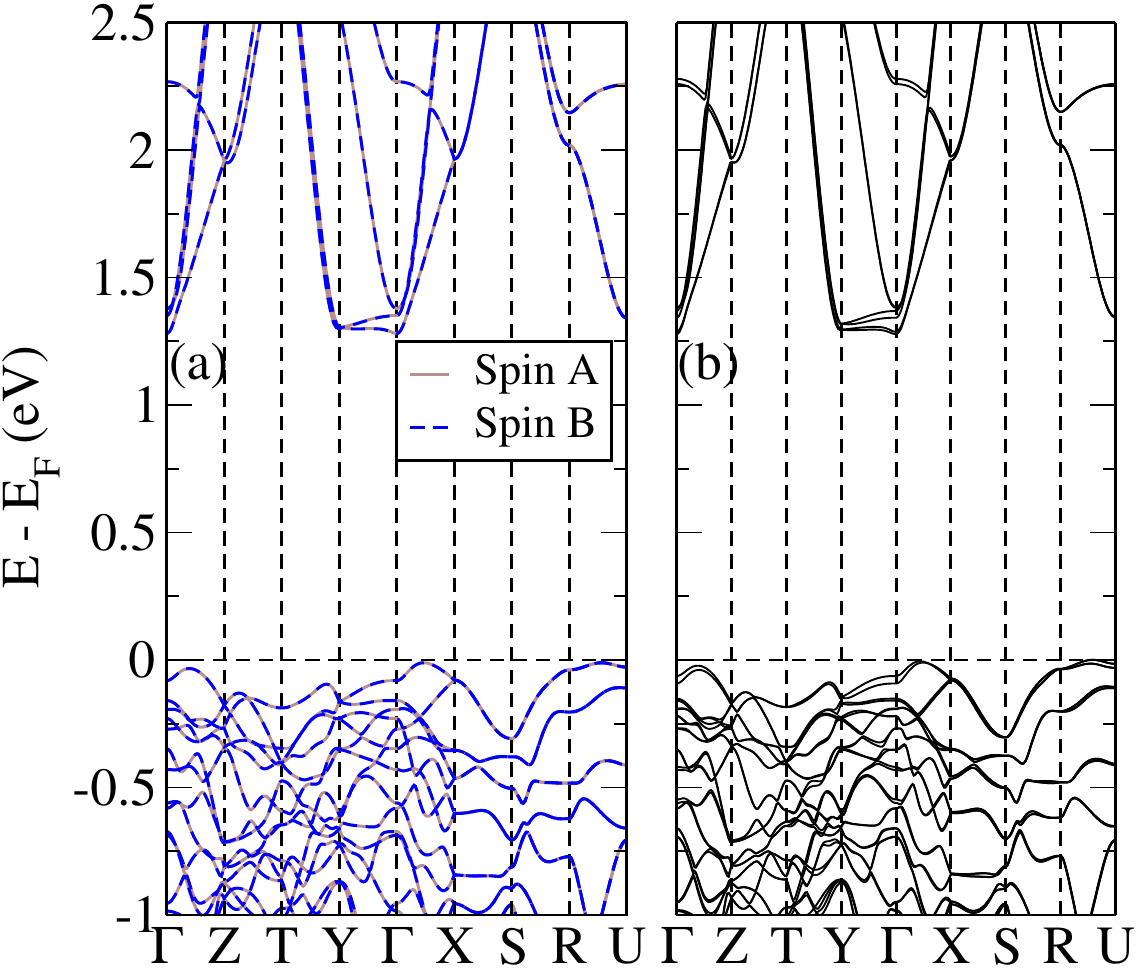}
	\caption{\label{Fig:CMOfullpath}Spin-polarized band dispersion for CMO from both spins marked as spin A and spin B is depicted in (a), while (b) represents the band dispersion considering spin-orbit interaction.}
\end{figure}

\begin{figure}
\includegraphics[scale=0.44]{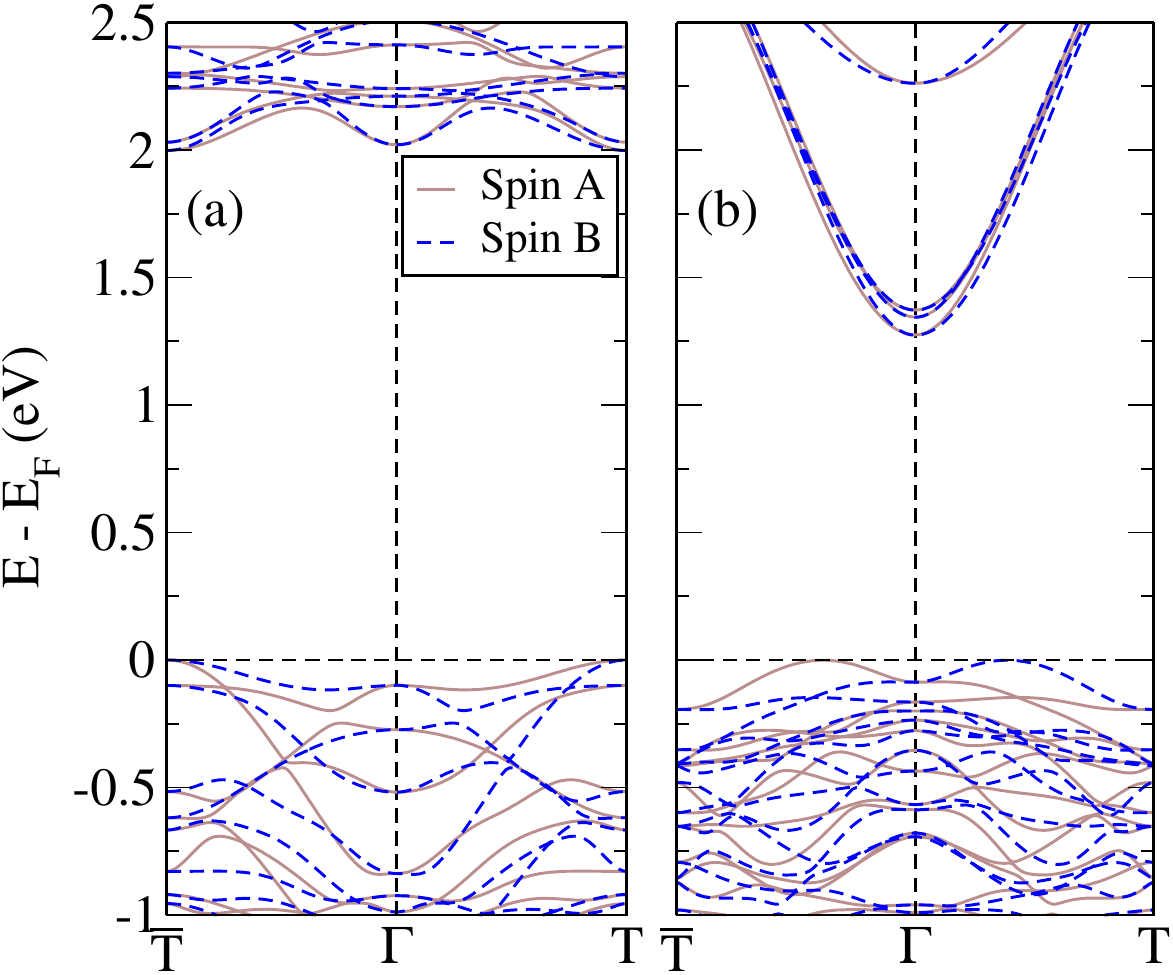}
	\caption{\label{Fig:BFOCMO_TGT}Spin-polarized band dispersion for BFO and CMO for both antiferromagnetic sublattices are displayed in (a) and (b), respectively, along the $\overline{\text{T}} \to \Gamma \to \text{T}$ direction.}
\end{figure}

The compounds BiFeO$_3$ (BFO) and CaMnO$_3$ (CMO) are found in nominal oxidation states Bi$^{3+}$Fe$^{3+}$O$_3^{2-}$ and Ca$^{2+}$Mn$^{4+}$O$_3^{2-}$, respectively, in their bulk forms. We start by comparing the total energies of different possible antiferromagnetic configurations, namely, A-type, C-type, and G-type antiferromagnets (AFMs) as depicted in the \cref{Fig:structure}(d),(e), and (f), which results in no net magnetic moment, alongside the ferromagnetic configuration. \cref{tab:Magnetic Configuration} shows the relative energies for different magnetic configurations, suggesting G-type AFM to have the lowest energy for both BFO and CMO. In G-type AFM, the magnetic moments are antiparallel for the nearest neighbors in all directions, as shown in \cref{Fig:structure}(f). Unless stated otherwise, the subsequent results regarding the unit cell correspond to the lowest energy G-type AFM configuration. However, CMO exhibits complex magnetic behavior at low temperatures, showing a paramagnet to AFM transition below the N\'{e}el temperature $T_N\sim$125~K, and a weak ferromagnetic behavior below $\sim$10~K in its ground state \cite{CMO_ordering_temp}. On the other hand BFO shows antiferromagnetic ordering below the N\'{e}el temperature $\sim$640~K \cite{BFOFerroPara}.

The band dispersion for both spin sublattices, marked as spin A and spin B for BFO and CMO, shown in \cref{Fig:BFOfullpath}(a) and \cref{Fig:CMOfullpath}(a), suggests an indirect band gap of $\sim$1.91~eV for BFO and $\sim$1.23~eV for CMO. We find projected magnetic moments of $\sim$4.16~$\mu_B$ and $\sim$2.74~$\mu_B$ at the Fe and Mn sites in BFO and CMO, respectively, suggesting a high spin configuration. We note that projected magnetic moments are usually underestimated within a plane wave basis set. The calculated band gap of CMO and the magnetic moment of the Mn atom are in good agreement with the experimentally reported range of $\sim$1.1-1.2~eV \cite{CMO_bandgap2} and $\sim$3~$\mu_B$ \cite{CMO_magmom}, respectively. On the other hand, due to limited experimental studies on orthorhombic BFO, the experimental band gap the magnetic moment of the Fe atom remains inaccessible. The band dispersions, shown in \cref{Fig:BFOfullpath}(a) and \cref{Fig:CMOfullpath}(a) for BFO and CMO, respectively, reveals a perfect overlap of both spins for both systems along the chosen high-symmetry directions. However, from a symmetry perspective, these compounds break the combined inversion $|$ time reversal ($\mathcal{PT}$) and combined spin rotation $|$ translation ($U\tau$) symmetries \cite{PhysRevB.102.014422}, as illustrated in \cref{Fig:structure}(c). As a consequence, the opposite spin bands cannot be degenerate everywhere in the Brillouin zone, as confirmed along the high-symmetry $\overline{\text{T}}\rightarrow\Gamma\rightarrow\text{T}$ direction for both compounds, displayed in \cref{Fig:BFOCMO_TGT}(a) and \cref{Fig:BFOCMO_TGT}(b) for BFO and CMO, respectively. Our observation is consistent with a recent theoretical prediction for CaMnO$_3$ as an altermagnet candidate \cite{Fernandes2024}.

%----------------------SOI---------------------------------------
%\subsection{Spin-orbit interaction}
%
\begin{table}
\caption{\label{tab:magnetic orientation_unitcell}The relative energies for the unit cell along different crystallographic directions including spin-orbit interaction in meV per formula unit are tabulated here.}
\begin{ruledtabular}
    \begin{tabular}{l..}
     Spin quantization axis & \multicolumn{1}{c}{BFO} & \multicolumn{1}{c}{CMO} \\
    \hline
      (100) & 0.265 & 0.055\\
      (010) & 0.497 & 0.017\\
      (001) & 0.0   &  0.0 \\
      (110) & 0.355 & 0.027\\
    \end{tabular}
\end{ruledtabular}
\end{table}
Upon considering the spin-orbit interaction (SOI) in our DFT calculations, we find an anisotropy in the spin-quantization axis based on the different crystallographic directions. Here, we calculate the relative energies for different spin quantization directions {\em viz.} (100), (010), (001), and (110) for both BFO and CMO, and tabulate the results in \cref{tab:magnetic orientation_unitcell}. The table reveals (001) as the preferred spin-quantization direction for both compounds. The band dispersions reveal minor changes upon considering SOI, as seen from \cref{Fig:BFOfullpath}(b) and \cref{Fig:CMOfullpath}(b).

%-----------------------Symmetry analysis-------------------------------------------

\subsection{\label{sec:Symmetry}Symmetry analysis}
\subsubsection{\label{subsec:SGsymmetry}Space Group Symmetry}
%----------------------NM----------------------------------------------
\begin{table*}
\caption{\label{tab:NMSpaceGroup}All the symmetry operations of the space group $Pnma$ are listed here.}
\begin{ruledtabular}
    \begin{tabular}{::}
    \{1|0\} : (x, y, z)\to (x, y, z) & \{2_{010}|0~ \frac{1}{2} ~0\} : (x, y, z)\to (-x, y+\frac{1}{2}, -z) \\
    \{2_{001}|\frac{1}{2}~ 0 ~\frac{1}{2}\}: (x, y, z)\to (-x+\frac{1}{2}, -y, z+\frac{1}{2}) & \{2_{100}|\frac{1}{2}~ \frac{1}{2} ~\frac{1}{2}\}: (x, y, z)\to (x+\frac{1}{2}, -y+\frac{1}{2}, -z+\frac{1}{2}) \\
    \{-1|0\}: (x, y, z)\to (-x, -y, -z) & \{m_{010}|0~ \frac{1}{2} ~0\}: (x, y, z)\to (x, -y+\frac{1}{2}, z) \\
    \{m_{001}|\frac{1}{2}~ 0 ~\frac{1}{2}\}: (x, y, z)\to (x+\frac{1}{2}, y, -z+\frac{1}{2}) & \{m_{100}|\frac{1}{2}~ \frac{1}{2} ~\frac{1}{2}\}: (x, y, z)\to (-x+\frac{1}{2}, y+\frac{1}{2}, z+\frac{1}{2})
    \end{tabular}
\end{ruledtabular}
\end{table*}

% ===== Pnm'a' =====
\begin{table*}
\caption{\label{tab:MSGSpaceGroup}All the symmetry operations of the magnetic space group $Pnm'a'$ are listed here.}
\begin{ruledtabular}
    \begin{tabular}{::}
    %\hline
    %\hline
    \multicolumn{1}{c}{$G_U$ (Unitary)} & \multicolumn{1}{c}{$G_{AU}$ (Antiunitary) = $\mathcal{T}(G-G_U)$} \\
    \hline
   \{1|0\} : (x, y, z) \to (x, y, z) & \mathcal{T} \{2_{001}|\frac{1}{2}~ 0 ~\frac{1}{2}\} : (x, y, z) \to (-x+\frac{1}{2}, -y, z+\frac{1}{2}) \\ 
   \{2_{100}|\frac{1}{2}~ \frac{1}{2} ~\frac{1}{2}\} : (x, y, z) \to (x+\frac{1}{2}, -y+\frac{1}{2}, -z+\frac{1}{2}) & \mathcal{T} \{2_{010}|0~ \frac{1}{2} ~0\} : (x, y, z)\to (-x, y+\frac{1}{2}, -z) \\
   \{-1|0\} : (x, y, z) \to (-x, -y, -z) & \mathcal{T} \{m_{001}|\frac{1}{2}~ 0 ~\frac{1}{2}\} : (x, y, z) \to (x+\frac{1}{2}, y, -z+\frac{1}{2}) \\
   \{m_{100}|\frac{1}{2}~\frac{1}{2} ~\frac{1}{2}\} : (x, y, z) \to (-x+\frac{1}{2}, y+\frac{1}{2}, z+\frac{1}{2}) & \mathcal{T} \{m_{010}|0~ \frac{1}{2} ~0\} : (x, y, z)\rightarrow (x, -y+\frac{1}{2}, z) \\
    \end{tabular}
\end{ruledtabular}
\end{table*}
%----------------------G-type-SOI----------------------------------------------
%\begin{table*}
%\caption{\label{tab:MSGSpaceGroupSOC}Different symmetry operations of the magnetic space group for G-type AFM with SOI: $Pn'ma'$.}
%\begin{ruledtabular}
%    \begin{tabular}{::}
    %\hline
    %\hline
%    \multicolumn{1}{c}{$G_U$ (Unitary)} & %\multicolumn{1}{c}{$G_{AU}$ (Antiunitary) = $\mathcal{T}(G-G_U)$} \\
%    \hline
%   \{1|0\} : (x, y, z) \to (x, y, z) & \mathcal{T} %\{2_{001}|\frac{1}{2}~ 0 ~\frac{1}{2}\} : (x, y, z) \to (-x+\frac{1}{2}, -y, z+\frac{1}{2}) \\ 
%   \{2_{010}|0~ \frac{1}{2} ~0\} : (x, y, z)\to (-x, y+\frac{1}{2}, -z) & \mathcal{T} \{2_{100}|\frac{1}{2}~ \frac{1}{2} ~\frac{1}{2}\} : (x, y, z) \to (x+\frac{1}{2}, -y+\frac{1}{2}, -z+\frac{1}{2}) \\
%   \{-1|0\} : (x, y, z) \to (-x, -y, -z) & \mathcal{T} \{m_{001}|\frac{1}{2}~ 0 ~\frac{1}{2}\} : (x, y, z) \to (x+\frac{1}{2}, y, -z+\frac{1}{2}) \\
%   \{m_{010}|0~ \frac{1}{2} ~0\} : (x, y, z)\rightarrow (x, -y+\frac{1}{2}, z) & \mathcal{T} \{m_{100}|\frac{1}{2}~\frac{1}{2} ~\frac{1}{2}\} : (x, y, z) \to (-x+\frac{1}{2}, y+\frac{1}{2}, z+\frac{1}{2}) \\
%    \end{tabular}
%\end{ruledtabular}
%\end{table*}

% ===== Pn'ma' =====
\begin{table*}
\caption{\label{tab:MSGSpaceGroupCtype}All the symmetry operations of the magnetic space group $Pn'ma'$ are listed here.}
\begin{ruledtabular}
    \begin{tabular}{::}
    \multicolumn{1}{c}{$G_U$ (Unitary)} & \multicolumn{1}{c}{$G_{AU}$ (Antiunitary) = $\mathcal{T}(G-G_U)$} \\
    \hline
   \{1|0\} : (x, y, z) \to (x, y, z) & \mathcal{T} \{2_{001}|\frac{1}{2}~ 0 ~\frac{1}{2}\} : (x, y, z) \to (-x+\frac{1}{2}, -y, z+\frac{1}{2}) \\ 
   \{2_{010}|0~ \frac{1}{2} ~0\} : (x, y, z) \to (-x, y+\frac{1}{2}, -z) & \mathcal{T} \{2_{100}|\frac{1}{2}~ \frac{1}{2} ~\frac{1}{2}\} : (x, y, z) \to (x+\frac{1}{2}, -y+\frac{1}{2}, -z+\frac{1}{2}) \\
   \{-1|0\} : (x, y, z) \to (-x, -y, -z) & \mathcal{T} \{m_{001}|\frac{1}{2}~ 0 ~\frac{1}{2}\} : (x, y, z) \to (x+\frac{1}{2}, y, -z+\frac{1}{2}) \\
   \{m_{010}|0~ \frac{1}{2} ~0\} : (x, y, z) \to (x, -y+\frac{1}{2}, z) & \mathcal{T} \{m_{100}|\frac{1}{2}~ \frac{1}{2} ~\frac{1}{2}\} : (x, y, z) \to (-x+\frac{1}{2}, y+\frac{1}{2}, z+\frac{1}{2}) \\
    \end{tabular}
\end{ruledtabular}
\end{table*}
% ===== Pn'm'a =====
\begin{table*}
\caption{\label{tab:MSGSpaceGroupAtype}All the symmetry operations of the magnetic space group $Pn'm'a$ are listed here.}
\begin{ruledtabular}
    \begin{tabular}{::}
    \multicolumn{1}{c}{$G_U$ (Unitary)} & \multicolumn{1}{c}{$G_{AU}$ (Antiunitary) = $\mathcal{T}(G-G_U)$} \\
    \hline
   \{1|0\} : (x, y, z) \to (x, y, z) & \mathcal{T} \{2_{010}|0~ \frac{1}{2} ~0\}: (x, y, z)\to (-x, y+\frac{1}{2}, -z) \\
   \{2_{001}|\frac{1}{2}~ 0 ~\frac{1}{2}\} : (x, y, z) \to (-x+\frac{1}{2}, -y, z+\frac{1}{2}) & \mathcal{T} \{2_{100}|\frac{1}{2}~ \frac{1}{2} ~\frac{1}{2}\} : (x, y, z) \to (x+\frac{1}{2}, -y+\frac{1}{2}, -z+\frac{1}{2}) \\
   \{-1|0\} : (x, y, z) \to (-x, -y, -z) & \mathcal{T} \{m_{010}|0~ \frac{1}{2} ~0\} : (x, y, z)\to (x, -y+\frac{1}{2}, z) \\
   \{m_{001}|\frac{1}{2}~ 0 ~\frac{1}{2}\} : (x, y, z) \to (x+\frac{1}{2}, y, -z+\frac{1}{2}) & \mathcal{T} \{m_{100}|\frac{1}{2}~\frac{1}{2} ~\frac{1}{2}\} : (x, y, z) \to (-x+\frac{1}{2}, y+\frac{1}{2}, z+\frac{1}{2}) \\
    \end{tabular}
\end{ruledtabular}
\end{table*}
% ===== =====
In this section, we discuss in detail the space group symmetries and magnetic space group (MSG) symmetries without spin-orbit interaction (SOI) for both compounds. BFO and CMO have the same orthorhombic symmetry as discussed in \cref{sec:method}; hence, all the symmetry analyses and conclusions drawn are equally applicable to both compounds. The $Pnma$ space group contains 8 unitary symmetry operations ($G \equiv G_U$) listed in \cref{tab:NMSpaceGroup}. Here, $1$ and $-1$ are the identity and the spatial inversion operation; $2_{100}, 2_{010}, 2_{001}$ are the two-fold ($\pi$) anticlockwise rotation about the $[100],[010],[001]$ axes, respectively; $m_{100}, m_{010}, m_{001}$ are the reflections in $[100],[010],[001]$ planes, respectively; vectors $(\frac{1}{2}~ \frac{1}{2} ~\frac{1}{2})$, $(0~ \frac{1}{2} ~0)$, and $(\frac{1}{2}~ 0 ~\frac{1}{2})$ are the nonprimitive translations. In this paper, we adopt all the notations similar to the Bilbao Crystallographic Server \cite{Bilbao2006, Aroyo:xo5013}.

Considering the magnetic arrangement of a system, the combined structural and magnetic symmetry may be described by a magnetic space group (MSG), where the time reversal operator $\mathcal{T}$ may combine with some of the group elements \cite{DresselhausGroupTheory08}. The MSGs without and with SOI are $Pnm'a'$ and $Pn'ma'$, respectively, for both compounds in the G-type antiferromagnetic configuration \cite{MSGbilbao2015, BradleySymmetrySolid}, with the symmetry operations listed in \cref{tab:MSGSpaceGroup} and \cref{tab:MSGSpaceGroupCtype}, respectively. Half of the elements of $G ~(Pnma)$ are included as unitary operations ($G_{U}$), while the rest are included as antiunitary operations $G_{AU}=\mathcal{T}(G-G_U)$ in the magnetic space groups.

%----------------Spin splitting behavior-----------------------------
\subsubsection{\label{subsec:spinsplitting}Spin-splitting without SOI}
Here, we discuss the splitting of the opposite-spin bands in different parts of the Brillouin zone (BZ) without considering SOI in our calculations. Although the band dispersions without SOI depicted in \cref{Fig:BFOfullpath}(a) and \cref{Fig:CMOfullpath}(a) for BFO and CMO, respectively, do not exhibit any spin-splitting feature, upon examining the band dispersion along the high-symmetry direction $\overline{\text{T}} (0, -\frac{\pi}{b}, \frac{\pi}{c})\to\Gamma\to\text{T} (0, \frac{\pi}{b}, \frac{\pi}{c})$, pronounced spin-splitting is observed from \cref{Fig:BFOCMO_TGT}(a) and \cref{Fig:BFOCMO_TGT}(b) for BFO and CMO, respectively. To further elucidate the spin-splitting, we have plotted the 3D band dispersion for a select pair of bands in different planes of the BZ. The spin degeneracy remains protected in the $k_z=0$ plane, as seen from the isoenergetic contour for $E - E_F = -0.5$~eV in the $k_x$-$k_y$ and the 3D band dispersion as a function of $(k_x, k_y)$, displayed in \cref{fig:BFO3D}(b) and \cref{fig:BFO3D}(c), respectively, for the pair of bands within a small energy range $[-0.6,-0.4]$~eV relative to the Fermi level, intersecting a horizontal red dashed line shown in \cref{fig:BFO3D}(a). Further, $k_y=0$ plane also preserves the spin degeneracy, as seen from isoenergetic contour for $E - E_F = -0.5$~eV displayed in \cref{fig:BFO3D}(d) for the same pair of bands. However, isoenergetic contour for $E - E_F = -0.5$~eV in the $k_y$-$k_z$ plane and the the 3D band dispersion as a function of ($k_y$, $k_z$) for $k_x=0$, shown in \cref{fig:BFO3D}(e) and \cref{fig:BFO3D}(f), respectively, for the same pair of bands exhibit a pronounced spin-splitting. The nature of the isoenergetic contour in the $k_y$-$k_z$ plane, as seen from \cref{fig:BFO3D}(e) unequivocally confirms the altermagnetic feature in BFO. Likewise, the 3D band dispersion and isoenergetic contour plots in various planes, depicted in \cref{fig:CMO3D}, also unveil altermagnetic features in CMO.

%-------------------MSG symmetry analysis-----------------------------------------------
\subsubsection{\label{subsec:MSGsymmetry}Magnetic space group symmetry and spin splitting}
%---------------------------------3D figure--------------------------------------------
\begin{figure*}
    \includegraphics[scale=0.65]{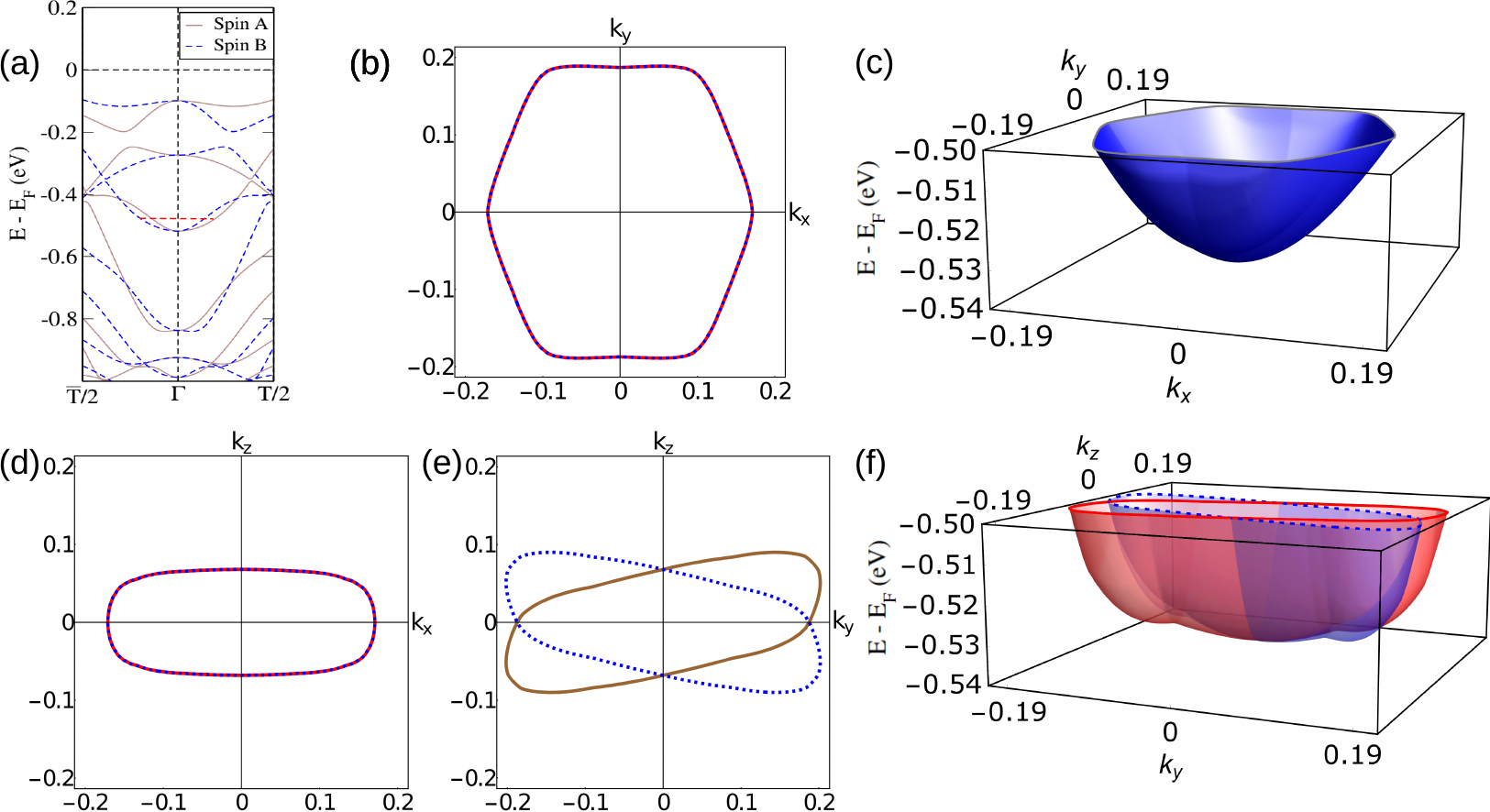}
    \caption{\label{fig:BFO3D}The electronic structure and spin degeneracy of BiFeO$_3$ is critically analyzed in this figure. Panel (a) shows the spin-polarized bands along $\overline{\text{T}}/2 \to \Gamma \to \text{T}/2$ direction. We choose a pair of spin-split bands in the valence band within an energy window $[-0.6, -0.4]$~eV intersecting a red dashed line at $E - E_F = -0.5$~eV. The isoenergetic contour in the $k_z = 0$ plane for $E - E_F = -0.5$~eV and the 3D band dispersion as a function of $(k_x, k_y)$ are depicted in (b) and (c), respectively. Panel (d) and (e) depict the isoenergetic contours of the same pair of bands in the $k_y = 0$ and $k_x = 0$ planes, respectively, for $E - E_F = -0.5$~eV, while the 3D dispersion as a function of $(k_y, k_z)$ with $k_x = 0$ is shown in (f).}
\end{figure*}
\begin{figure*}
    \includegraphics[scale=0.65]{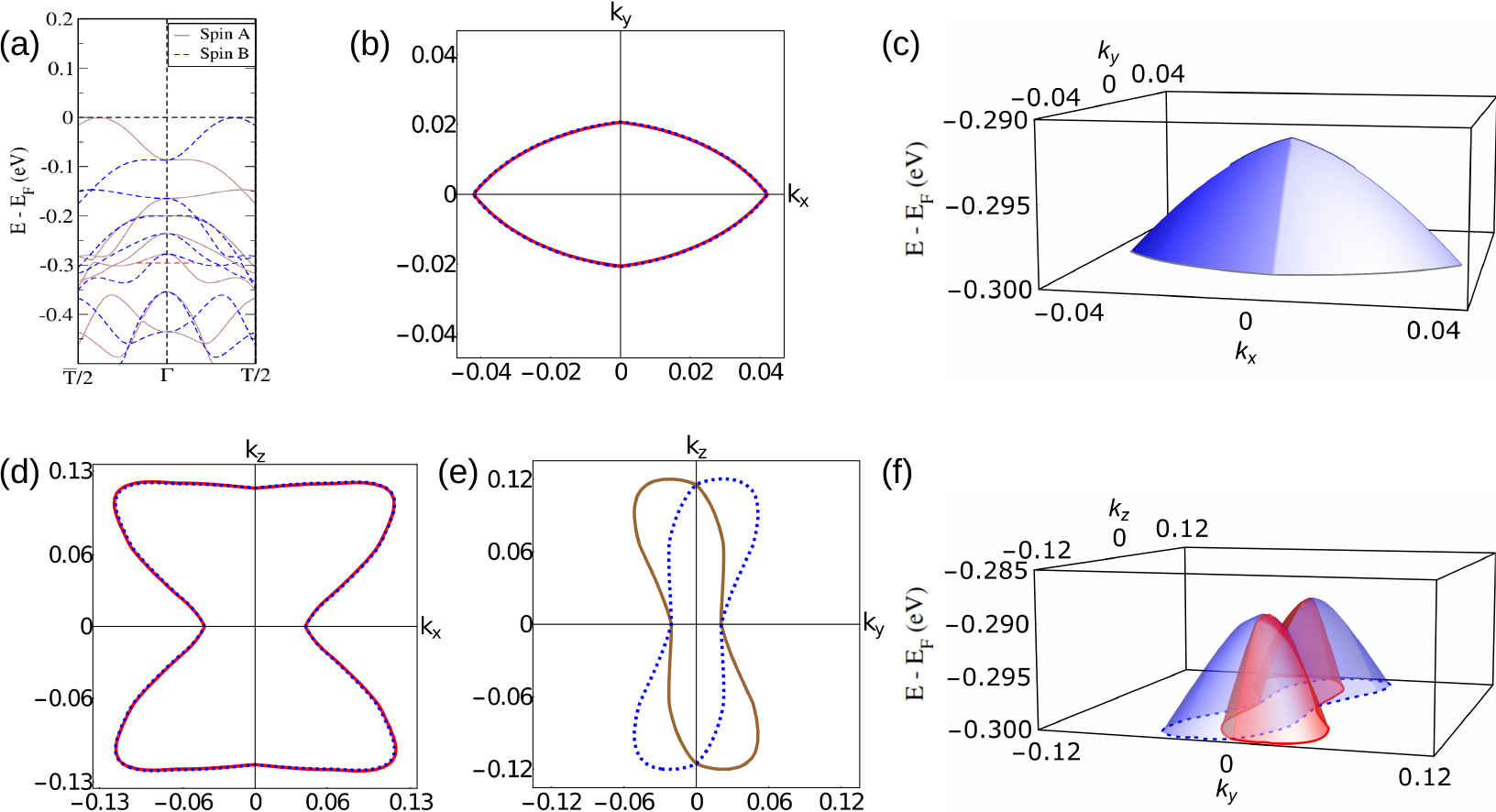}
    \caption{\label{fig:CMO3D}The electronic structure and spin degeneracy of CaMnO$_3$ is critically analyzed in this figure. Panel (a) shows the spin-polarized bands along $\overline{\text{T}}/2 \to \Gamma \to \text{T}/2$ direction. We choose a pair of spin-split bands in the valence band intersecting a red dashed line at $E - E_F = -0.3$~eV. The isoenergetic contour in the $k_z = 0$ plane for $E - E_F = -0.3$~eV and the 3D band dispersion as a function of $(k_x, k_y)$ are depicted in (b) and (c), respectively. Panel (d) and (e) depict the isoenergetic contours of the same pair of bands in the $k_y = 0$ and $k_x = 0$ planes, respectively, for $E - E_F = -0.3$~eV, while the 3D dispersion as a function of $(k_y, k_z)$ with $k_x = 0$ is shown in (f).}
\end{figure*}
%------------------------------------------------------------------------------------------
\begin{figure*}
    \includegraphics[scale=0.65]{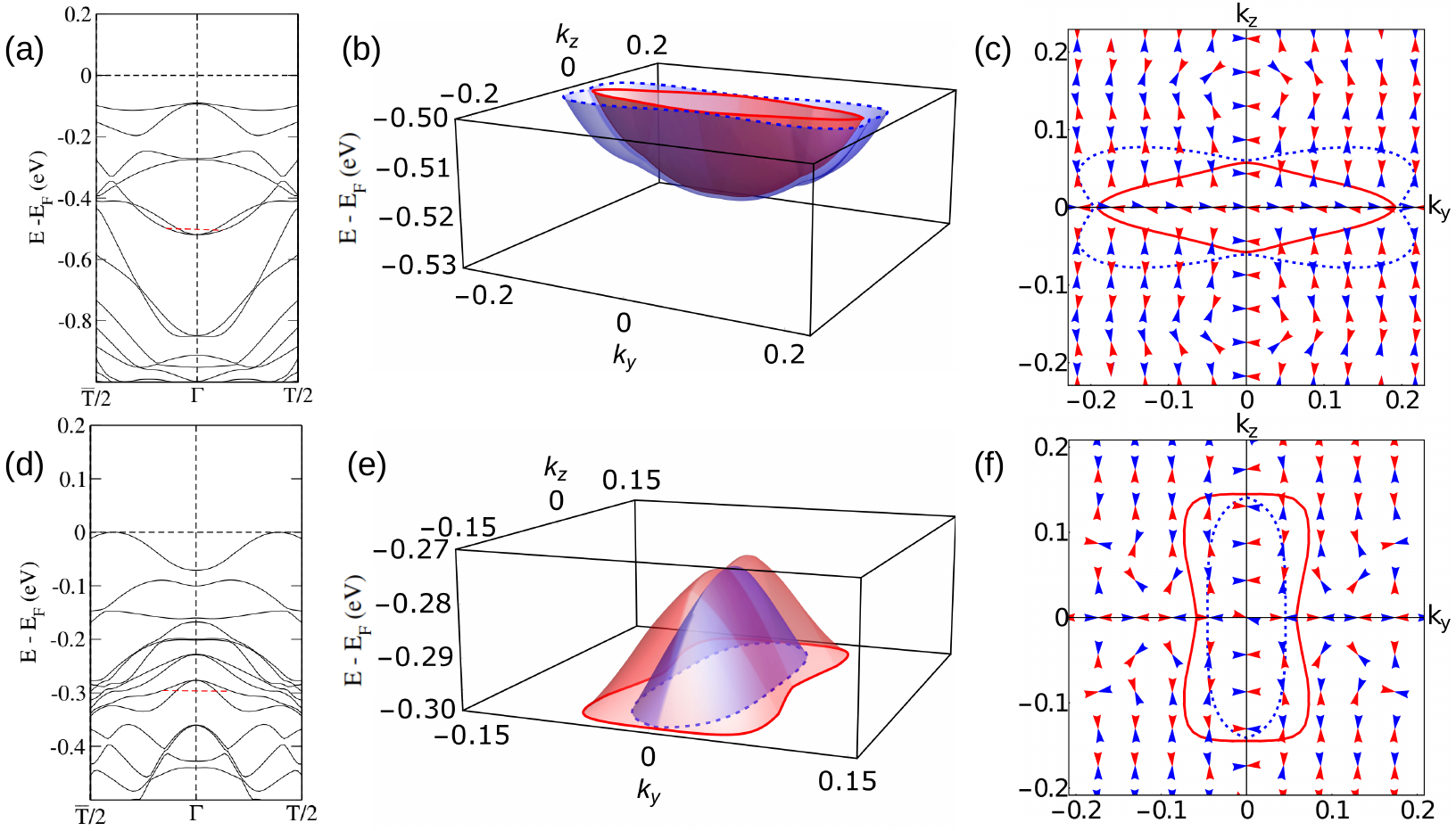}
    \caption{\label{fig:socbfocmo}The electronic structure of BiFeO$_3$ and CaMnO$_3$ are critically analyzed within spin-orbit interaction. Panels (a) and (d) show the band structure of BiFeO$_3$ and CaMnO$_3$, respectively, along $\overline{\text{T}}/2 \to \Gamma \to \text{T}/2$ direction. For a chosen pair of bands at $E - E_F = -0.5$~eV for BiFeO$_3$ and $E - E_F = -0.3$~eV for CaMnO$_3$ (marked with a horizontal red dashed line in panels (a) and (d)), the energy dispersions are shown as a function of ($k_y, k_z$) (3D bands) in panel (b) and (e). Panels (c) and (f) exhibit the isoenergetic contours for the same pair of bands at $E - E_F = -0.5$~eV and $E - E_F = -0.3$~eV with projected spins for BiFeO$_3$ and CaMnO$_3$, respectively.}
\end{figure*}
%------------------------------------------------------------
The group of wave vectors at the high-symmetry points in the BZ shown in \cref{Fig:structure} (b) preserve all the symmetry operations of the MSG $Pnm'a'$ that contain both unitary and antiunitary operations. The antiunitary operations flip the spin, keeping the coordinate of the high-symmetry points invariant and imposing the spin degeneracy, as evident from \cref{Fig:BFOfullpath}(a) and \cref{Fig:CMOfullpath}(a) for BFO and CMO, respectively. $(u, v, 0)$ with $(u, v) \in ([-\pi/a, \pi/a], [-\pi/b, \pi/b])$ represents a generic $\vec{k}$-vector in the $k_x$-$k_y$ plane. The symmetry operations in the MSG $Pnm'a'$ that leave a generic $\vec{k}$-vector in the $k_x$-$k_y$ plane invariant are
\begin{align}
    \{1|0\} &: (u,v,0) \to (u,v,0) \text{ and} \nonumber \\
    \mathcal{T} \left\{2_{001} \middle| \frac{1}{2}~ 0 ~\frac{1}{2} \right\} &: (u,v,0) \to (u,v,0), \label{eq:kxky}
\end{align}
as seen from \cref{tab:MSGSpaceGroup}. The antiunitary operation $\mathcal{T}\{2_{001}|\frac{1}{2}~ 0 ~\frac{1}{2}\}$ that connects both magnetic sublattices ensures the spin degeneracy throughout the entire $k_x$-$k_y$ plane for both compounds, as seen from \cref{fig:BFO3D}(b),(c) and \cref{fig:CMO3D}(b),(c).

Similarly, considering $(u, 0, w)$ as a generic coordinate of a $\vec{k}$ in the $k_x$-$k_z$ plane, we find the symmetry operations of the MSG $Pnm'a'$ that keep it invariant are (see \cref{tab:MSGSpaceGroup})
\begin{align}
    \{1|0\} &: (u,0,w) \to (u,0,w) \text{ and} \nonumber \\
    \mathcal{T} \left\{ 2_{010} \middle| 0~ \frac{1}{2} ~0 \right\} &: (u,0,w) \to (u,0,w). \label{Eq:kxkz}
\end{align}
The antiunitary operation $\mathcal{T}\{2_{010}|0~ \frac{1}{2} ~0\} : (u,0,w)$ connecting both spin sublattices preserves a spin degeneracy in this plane, as confirmed from the isoenergetic contours in \cref{fig:BFO3D}(d) and \cref{fig:CMO3D}(d).

When we consider the $k_y$-$k_z$ plane, a generic $\vec{k}$-vector is represented with the coordinates $(0, v, w)$. The symmetry operations in the MSG $Pnm'a'$ that keep the $\vec{k}$-vector invariant are (see \cref{tab:MSGSpaceGroup})
\begin{align}
    \{1|0\} &: (0, v, w) \to (0, v, w) \text{ and} \nonumber \\
    \left\{ m_{100} \middle| \frac{1}{2}~ \frac{1}{2} ~\frac{1}{2} \right\} &: (0, v, w) \to (0, v, w). \label{Eq:kykz}
\end{align}
None of these symmetry operations connect the magnetic sublattices with opposite spins, leading to broken spin degeneracy of the bands in this plane, as evident from \cref{fig:BFO3D}(e),(f) and \cref{fig:CMO3D}(e),(f), except for the two nodes at $k_y = 0$ and the other two nodes at $k_z = 0$ (see isoenergetic contour in \cref{fig:BFO3D}(e) for BFO and \cref{fig:CMO3D}(e) for CMO). The $\vec{k}$-vectors at the nodes at $k_y = 0$ and $k_z = 0$ may be represented by generic coordinates $(0, 0, w)$ and $(0, v, 0)$, respectively, which belong to the two spin degenerate $k_x$-$k_z$ and $k_x$-$k_y$ planes, respectively. Thus, our previous argument explains how the spin degeneracy is protected at these nodes.

%---------------------Other MSGs----------------------------------------------------------
\subsection{\label{sec:GeneralDescription}Prediction of spin-splitting for other MSGs}
Here we explore the possibility of spin-splitting in two other possible MSGs within the $Pnma$ space group without considering SOI. We begin by elaborating on the other two magnetic structures, A and C-type, for the unit cell. For A-type and C-type antiferromagnetic ordering, illustrated in \cref{Fig:structure}(d) and \cref{Fig:structure}(e), respectively, the MSG without SOI becomes $Pn'm'a$ and $Pn'ma'$, respectively \cite{Bilbao2006, Aroyo:xo5013, MagneticSpaceGroup}. The explicit symmetry operations are listed in \cref{tab:MSGSpaceGroupAtype} and \cref{tab:MSGSpaceGroupCtype}, respectively.
% ===== Table: A-type and C-type =====
\begin{table}
    \caption{\label{tab:antiUnitaryPlanes}The antiunitary operations belonging to the magnetic space group that keep the $\vec{k}$-vector invariant in different planes of the Brillouin zone are listed here.}
    \begin{ruledtabular}
        \begin{tabular}{cc*}
             MSG & BZ plane & \multicolumn{1}{c}{Antiunitary symmetry operation} \\
             \hline
              & $k_x = 0$ & \multicolumn{1}{c}{None, spin-splitting} \\
             $Pnm'a'$ & $k_y = 0$ & \mathcal{T} \{ 2_{010} | 0~\frac{1}{2}~0 \} : (u, 0, w) \to (u, 0, w) \\
              & $k_z = 0$ & \mathcal{T} \{ 2_{001} | \frac{1}{2}~0~\frac{1}{2} \} : (u, v, 0) \to (u, v, 0) \\
             \hline
              & $k_x = 0$ & \mathcal{T} \{ 2_{100}|\frac{1}{2}~\frac{1}{2}~\frac{1}{2} \} : (0, v, w) \to (0, v, w) \\
             $Pn'm'a$ & $k_y = 0$ & \mathcal{T} \{ 2_{010}|0~\frac{1}{2}~0 \} : (u, 0, w) \to (u, 0, w) \\
              & $k_z = 0$ & \multicolumn{1}{c}{None, spin-splitting} \\
            \hline
              & $k_x = 0$ & \mathcal{T} \{ 2_{100} | \frac{1}{2}~\frac{1}{2}~\frac{1}{2} \} : (0, v, w) \to (0, v, w) \\
             $Pn'ma'$ & $k_y = 0$ & \multicolumn{1}{c}{None, spin-splitting} \\
              & $k_z = 0$ & \mathcal{T} \{ 2_{001} | \frac{1}{2}~0~\frac{1}{2} \} : (u, v, 0) \to (u, v, 0)
        \end{tabular}
    \end{ruledtabular}
\end{table}

%===============================================
The high-symmetry points in the BZ exhibit all the unitary and antiunitary symmetry operations of the MSGs $Pn'm'a$ and $Pn'ma'$, protecting the spin degeneracy at all high-symmetry points. Similar to our previous symmetry analysis, here we find that a generic $\vec{k}$-vector in the $k_x$-$k_y$ plane remains invariant under the symmetry operations $\{1|0\}$ and $\{m_{001}|\frac{1}{2}~ 0 ~\frac{1}{2}\}$ belonging to the MSG $Pn'm'a$, and the symmetry operations $\{1|0\}$ and $\mathcal{T} \{2_{001}| \frac{1}{2}~0~\frac{1}{2}\}$ belonging to the MSG $Pn'ma'$. Thus, although the spin degeneracy should be protected in the $k_x$-$k_y$ plane for a C-type antiferromagnet, a spin splitting should be observed in the same plane for an A-type antiferromagnet. \Cref{tab:antiUnitaryPlanes} tabulates the possible antiunitary symmetry operations for different planes in the BZ for the magnetic space groups $Pnm'a'$, $Pn'm'a$, and $Pn'ma'$, revealing the planes where no antiunitary symmetry operation protects band degeneracy of the opposite spin sublattices, leading to spin-splitting. As summarized in \cref{tab:antiUnitaryPlanes}, the BZ planes $k_x = 0$, $k_y = 0$, and $k_z = 0$ host spin-splitting for magnetic space groups $Pnm'a'$, $Pn'm'a$, and $Pn'ma'$, corresponding to G-type, A-type, and C-type antiferromagnetic configurations, respectively, where the first one has been verified from our DFT calculations.

%-----------------------------Hamiltonian character tables-----------------------------------------------
\begin{table*}
\caption{\label{tab:Hamiltonian}The transformation properties and the irreducible tensor up to the second order are listed according to the generator of the group of wave vector at $\Gamma$ point for $Pnm'a'$ MSG.}
\begin{ruledtabular}
    \begin{tabular}{ll....}
    Symmetrized matrix &  \multicolumn{1}{c}{Irreducible tensor} &  \multicolumn{1}{c}{$\{2_{100}|\frac{1}{2}~ \frac{1}{2} ~\frac{1}{2}\}$} & \multicolumn{1}{c}{$\{-1|0\}$}  & \multicolumn{1}{c}{$\mathcal{T}\{2_{001}|\frac{1}{2}~ 0 ~\frac{1}{2}\}$}\\
   \hline
    $\sigma_0$ & $k^2_x, k^2_y, k^2_z$ & 1 & 1 & 1 \\
      -  & $k_xk_y$ & -1 & 1 & 1\\
      -  & $k_xk_z$ & -1 & 1 & -1\\
    $\sigma_z$  & $k_yk_z$ & 1 & 1 & -1\\ 
    \end{tabular}
\end{ruledtabular}
\end{table*}

\begin{table*}
\caption{\label{tab:HamiltonianAtype}The transformation properties and the irreducible tensor up to the second order are listed according to the generator of the group of wave vector at $\Gamma$ point for $Pn'm'a$ MSG.}
\begin{ruledtabular}
    \begin{tabular}{ll....}
    Symmetrized matrix &  \multicolumn{1}{c}{Irreducible tensor} &  \multicolumn{1}{c}{$\{2_{001}|\frac{1}{2}~ 0 ~\frac{1}{2}\}$} & \multicolumn{1}{c}{$\{-1|0\}$}  & \multicolumn{1}{c}{$\mathcal{T}\{2_{100}|\frac{1}{2}~ \frac{1}{2} ~\frac{1}{2}\}$}\\
   \hline
    $\sigma_0$ & $k^2_x, k^2_y, k^2_z$ & 1 & 1 & 1 \\
    $\sigma_z$  & $k_xk_y$ & 1 & 1 & -1\\
      -  & $k_xk_z$ & -1 & 1 & -1\\
      -  & $k_yk_z$ & -1 & 1 & 1\\ 
    \end{tabular}
\end{ruledtabular}
\end{table*}

\begin{table*}
\caption{\label{tab:HamiltonianCtype}The transformation properties and the irreducible tensor up to the second order are listed according to the generator of the group of wave vector at $\Gamma$ point for $Pn'ma'$ MSG.}
\begin{ruledtabular}
    \begin{tabular}{ll....}
    Symmetrized matrix &  \multicolumn{1}{c}{Irreducible tensor} &  \multicolumn{1}{c}{$\{2_{010}|0~ \frac{1}{2} ~0\}$} & \multicolumn{1}{c}{$\{-1|0\}$}  & \multicolumn{1}{c}{$\mathcal{T}\{2_{100}|\frac{1}{2}~ \frac{1}{2} ~\frac{1}{2}\}$}\\
   \hline
    $\sigma_0$ & $k^2_x, k^2_y, k^2_z$ & 1 & 1 & 1 \\
      -  & $k_xk_y$ & -1 & 1 & -1\\
    $\sigma_z$  & $k_xk_z$ & 1 & 1 & -1\\
      -  & $k_yk_z$ & -1 & 1 & 1\\ 
    \end{tabular}
\end{ruledtabular}
\end{table*}
%---------------------------------------------------------------------------------------------

% ========== Hamiltonian ==========
\subsection{\label{subsec:Hamiltonian}Effctive two-band model Hamiltonian at $\Gamma$ point without SOI}
After identifying the symmetry operations that enforce the spin degeneracy in some parts of the BZ, below we deduce a symmetry-adapted Hamiltonian that allows spin splitting in the remaining part of the BZ.

\subsubsection{\label{subsec:HamiltonianG}Effective two-band model Hamiltonian for the MSG $Pnm'a'$}
First we construct the model Hamiltonian of the G-type antiferromagnet belonging to the $Pnm'a'$ MSG. Our approach involves defining two highly localized spin basis states at the magnetic sites according to the G-type antiferromagnetic ordering for both compounds. Subsequently, the effective two-band model Hamiltonian at a specific $\vec{k}$ is derived by incorporating the symmetry constraints imposed by the group of wave vector on the basis. In this context, the group of wave vector at $\Gamma$ point satisfies all the symmetry operations of the magnetic space group ($Pnm'a'$) without spin-orbit interaction (SOI), as listed in \cref{tab:MSGSpaceGroup}. By selectively focusing on those symmetry operations associated with the generators of the group, we present the representations and transformation properties of both the Pauli vector $\vec{\sigma}$ and the wave vector $\vec{k}$.

The generator of the MSG $Pnm'a'$ contains two unitary operations $\{2_{100}|\frac{1}{2}~\frac{1}{2} ~\frac{1}{2}\}$, $\{-1|0\}$ and one antiunitary operation $\mathcal{T}\{2_{001}|\frac{1}{2}~ 0 ~\frac{1}{2}\}$. The transformation properties of the Pauli matrix and the irreducible tensor operator $\vec{k}$ upto the second order under these generators are listed in \cref{tab:Hamiltonian}. Analyzing the \cref{tab:Hamiltonian}, the only possible invariant term in the Hamiltonian is found to be $\sigma_z k_yk_z$, hinting a quadratic in $\vec{k}$ nature of the nonrelativistic spin-splitting in the $k_y$-$k_z$ plane, consistent with the Ref. \cite{PhysRevB.102.014422}. Hence, the Hamiltonian takes the form
\begin{equation}
   H = H_0 + \alpha \sigma_z k_y k_z, \label{Eq:HamiltonianG}  
\end{equation}
where $H_0$ represents the Hamiltonian without considering the spin-splitting that may be represented by a standard tight-binding Hamiltonian, and $\alpha$ is a constant coefficient. The eigenvalues of the Hamiltonian mentioned in \cref{Eq:HamiltonianG} are $\varepsilon_0(\vec{k}) \pm \alpha k_y k_z$, with $\varepsilon_0(\vec{k})$ representing the band dispersion without the spin-splitting, which qualitatively explain the obtained results shown in the \cref{fig:BFO3D}(e),(f) and \cref{fig:CMO3D}(e),(f) for both compounds.

%---------Hamiltonian for A (C) type AFM-----------------------
\subsubsection{\label{subsec:HamiltonianAC}Effective two-band model Hamiltonian for the MSGs $Pn'm'a$ and $Pn'ma'$}
We extend our Hamiltonian formulation to the high-symmetry $\Gamma$ point for two predicted MSGs: $Pn'm'a$ and $Pn'ma'$, corresponding to A and C-type AFM structures, respectively. Following an approach similar to the above, we define two localized spin basis states at the magnetic sites according to the A and C-type AFM ordering. Then, the effective two-band model Hamiltonian at the $\Gamma$-point can be constructed by considering the symmetry constraints imposed by the group of wave vector on the basis. The group of wave vector at $\Gamma$ point maintain all the symmetries of the MSGs $Pn'm'a$ and $Pn'ma'$. From all the symmetries of both groups $Pn'm'a$ and $Pn'ma'$ listed in \cref{tab:MSGSpaceGroupAtype} and \cref{tab:MSGSpaceGroupCtype}, we consider only the generators of the group to determine the representation and transformation properties of the Pauli vector $\vec{\sigma}$ and the wave vector $\vec{k}$. The generators and the transformation properties of $\vec{\sigma}$ and $\vec{k}$ are presented in the \cref{tab:HamiltonianAtype} and \cref{tab:HamiltonianCtype} for the MSG $Pn'm'a$ and $Pn'ma'$. By analyzing these tables, we find that the only invariant terms that could exist in the Hamiltonian are $\sigma_z k_x k_y$ for MSG $Pn'm'a$ and $\sigma_z k_x k_z$ for MSG $Pn'ma'$. 

%========================================================
\begin{table}
    \caption{\label{tab:wyckoffMGS}The magnetic Wyckoff positions in the space group $Pnma$ for BiFeO$_3$ (CaMnO$_3$) are listed here.}
    \begin{ruledtabular}
        \begin{tabular}{lll}
            Site & Wyckoff & Coordinates \\
            & position & \\
             \hline
             &  & {($x,\frac{1}{4},z|0,m_y,0$)} \\
             Bi (Ca) & $4c$ & {($-x,\frac{3}{4},-z|0,m_y,0)$} \\
              &  & {($x+\frac{1}{2},\frac{1}{4},-z+\frac{1}{2}|0,m_y,0)$} \\
              &  & {($-x+\frac{1}{2},\frac{3}{4},z+\frac{1}{2}|0,m_y,0)$} \\
             \hline
              &  & {($0,0,0|m_x,m_y,m_z$)} \\
             Fe (Mn) & $4a$ & {($0,\frac{1}{2},0|-m_x,m_y,-m_z)$} \\
              &  & {($\frac{1}{2},\frac{1}{2},\frac{1}{2}|-m_x,m_y,m_z)$} \\
              &  & {($\frac{1}{2},0,\frac{1}{2}|m_x,m_y,-m_z)$} \\
            \hline
              &  & {($x,\frac{1}{4},z|0,m_y,0$)} \\
              O & $4c$ & {($-x,\frac{3}{4},-z|0,m_y,0)$} \\
              &  & {($x+\frac{1}{2},\frac{1}{4},-z+\frac{1}{2}|0,m_y,0)$} \\
              &  & {($-x+\frac{1}{2},\frac{3}{4},z+\frac{1}{2}|0,m_y,0)$} \\
              \hline
              &  & {($x,y,z|m_x,m_y,m_z$)} \\
              &  & {($-x,y+\frac{1}{2},-z|-m_x,m_y,-m_z$)} \\
              &  & {($-x,-y,-z|m_x,m_y,m_z$)} \\
             O & $8d$ & {($x,-y+\frac{1}{2},z|-m_x,m_y,-m_z$)}\\
              &  & {($x+\frac{1}{2},-y+\frac{1}{2},-z+\frac{1}{2}|-m_x,m_y,m_z$)} \\
              &  & {($-x+\frac{1}{2},-y,z+\frac{1}{2}|m_x,m_y,-m_z$)} \\
              &  & {($-x+\frac{1}{2},y+\frac{1}{2},z+\frac{1}{2}|-m_x,m_y,m_z$)} \\
              &  & {($x+\frac{1}{2},y,-z+\frac{1}{2}|m_x,m_y,-m_z$)} \\
        \end{tabular}
    \end{ruledtabular}
\end{table}
%====================================================

\subsection{\label{sec:SOI}Altermagnetism within spin-orbit interaction}
After critically analyzing the magnetic space group symmetries and finding a model Hamiltonian that remains invariant under them, we examine the influence of spin-orbit interaction on the physical properties of both altermagnetic compounds. The band dispersion for BFO changes slightly upon incorporating SOI, as evident from comparing \cref{Fig:BFOCMO_TGT}(a) and \cref{fig:socbfocmo}(a), while the band dispersion for CMO shows larger changes under SOI, as seen by comparing \cref{Fig:BFOCMO_TGT}(b) and \cref{fig:socbfocmo}(d). To closely understand the electronic structure, we consider a spin-split pair of bands for BFO and CMO marked with red dashed lines in \cref{fig:socbfocmo}(a) and \cref{fig:socbfocmo}(d), respectively, and plot the 3D band dispersion as a function of $(k_y, k_z)$ ($\varepsilon(k_x = 0, k_y, k_z)$) for both compounds, as shown in \cref{fig:socbfocmo}(b) and \cref{fig:socbfocmo}(e). \cref{fig:socbfocmo}(c) and \cref{fig:socbfocmo}(f) exhibit isoenergetic contours for the same pair of bands at $E-E_F=-0.5$~eV and $E-E_F=-0.3$~eV for BFO and CMO, respectively, with projected spin textures directly obtained from our DFT calculations.

Taking SOI into consideration, the magnetic space group (MSG) of both compounds is $Pn'ma'$, with corresponding magnetic point group (MPG) $m'mm'$ consisting of two orthogonal time-reversed mirror planes $\mathcal{T}m_{100}~\text{and}~\mathcal{T}m_{001}$, and a mirror plane $m_{010}$ without time-reversal (see \cref{tab:MSGSpaceGroupCtype}). These three orthogonal mirror planes allow only a net magnetic moment along the $[010]$ direction, as seen from \cref{tab:wyckoffMGS} that lists the magnetic Wyckoff positions and the allowed magnetic moments. Our DFT calculations reveal that both BFO and CMO have the major magnetic moment components along the easy axis $[001]$ that cancel due to an antiferromagnetic arrangement, leaving only small net moments of 0.036~$\mu_B$ and 0.287~$\mu_B$ for BFO and CMO, respectively, along the $[010]$ direction, consistent with the symmetry analysis, suggesting a weak ferromagnetism in both systems. Thus, SOI leads to a weak ferromagnetism in the altermagnetic BFO and CMO, requiring a noncollinear spin arrangement. The spin textures shown in \cref{fig:socbfocmo}(c) and \cref{fig:socbfocmo}(f) for BFO and CMO, respectively, also reveal a noncollinear nature. Thus, we find BFO and CMO to have mixed altermagnet-weak ferromagnet phases with the easy axis along $[001]$ direction.
% ===== Figure: AHC =====
\begin{figure}
\includegraphics[scale=0.52]{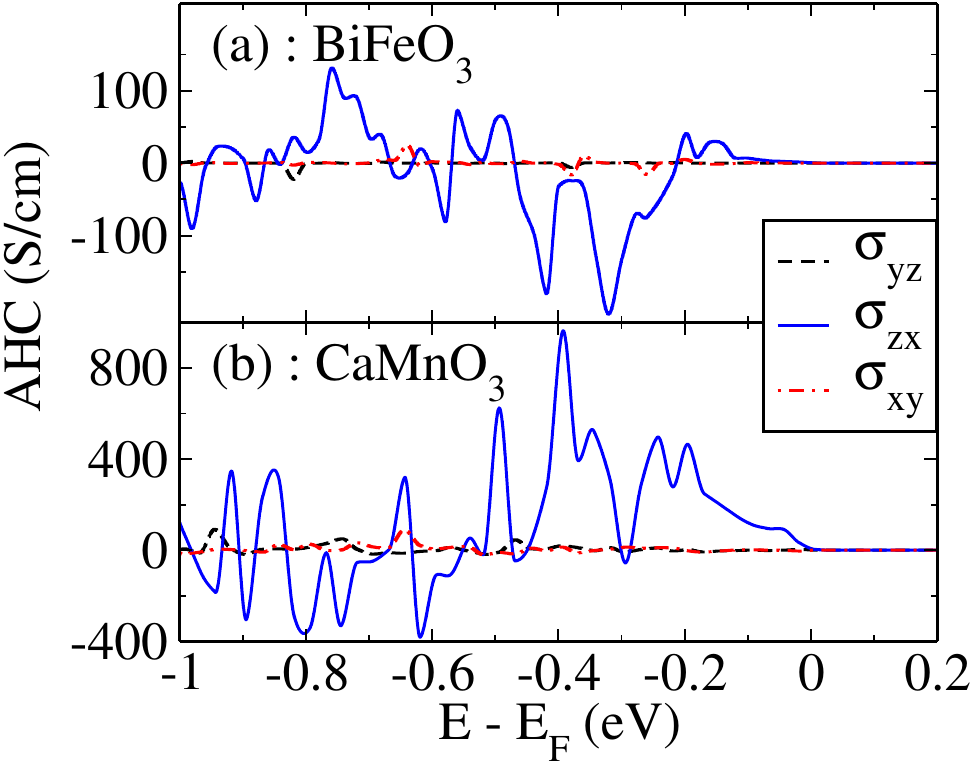}
	\caption{\label{Fig:ahc}Panels (a) and (b) show the variation of the anomalous Hall conductivity tensor components with energy for BiFeO$_3$ and CaMnO$_3$, respectively, upon considering spin-orbit interaction.}
\end{figure}
% ===== =====
The discussion on altermagnetism-induced weak ferromagnetism motivates us to investigate the anomalous Hall effect (AHE) in both compounds as the weak ferromagnetic component in an altermagnet often leads to a non-zero anomalous Hall conductivity \cite{SmejkalPRX22, Smejkal2021AHE}. Since the net magnetic moment is along the $y$ direction, one could expect only a significant value of $\sigma_{zx}$ in the anomalous Hall conductivity tensor, with the other two independent components, $\sigma_{xy}$ and $\sigma_{yz}$ to be vanishingly small. Our results, shown in \cref{Fig:ahc}(a) and \cref{Fig:ahc}(b) for BFO and CMO, respectively, resonate with our expectations, revealing a large $\sigma_{zx}$ value, comparable to that reported in Ref.~\cite{FakhredinePRB23}, and negligible $\sigma_{xy}$ and $\sigma_{yz}$ values for both compounds.

%\alert{CMO - MAGMOM- 0.2870 $\mu_B$  (total) or 0.07175 $\mu_B$  per formula unit along y-axis. AHC - (-0.4917 along x, 13.7425 along y, 0.5753 along z in the unit of S/cm) BFO -  MAGMOM - -0.0360 $\mu_B$  (total) or 0.009 $\mu_B$  per formula unit along y-axis. AHC - (0.0 along x, 0.0 along y, 0.0 along z in the unit of S/cm) }
%==============================================================
%\subsection{\label{subsec:HamiltonianGSOI}\add{Low energy model Hamiltonian for the G-type AFM with SOI:} $Pn'ma'$}

% ========== CONCLUSION ==========
\section{\label{sec:discussion}Conclusion}
To conclude our work, we studied bulk orthorhombic BFO and CMO within the first-principle density functional theory to understand its electronic structure and antiferromagnetic spin-splitting. Beginning with a chosen exchange-correlation functional and an appropriate Hubbard correction, our investigation unveiled an insulating state in both compounds characterized by a preferred antiferromagnetic order. We observe a notable spin-splitting phenomenon within the lowest energy antiferromagnetic configuration for both systems, particularly evident in the $k_y$-$k_z$ plane while maintaining spin degeneracy in other planes. Our critical examination of the spin-splitting features without spin-orbit interaction via conventional band dispersion, 3D bands, and isoenergetic contours obtained from our DFT calculations confirms these two compounds as altermagnets. We also present an analytical explanation of our DFT results by analyzing magnetic symmetry in detail and an insightful analytical model Hamiltonian based on the magnetic space group analysis. Additionally, we explore two other magnetic space groups corresponding to A-type and C-type antiferromagnetic arrangements within the $Pnma$ space group that may host the spin-splitting in different Brillouin zone planes and their probable form of the Hamiltonian. We further present a comprehensive analysis of the electronic structure incorporating spin-orbit interaction via usual band dispersion, 3D band dispersion, isoenergetic contours, and projected spin-textures derived from our DFT calculations. Our detailed examination, guided by magnetic symmetry analysis, reveals the presence of weak ferromagnetism in both compounds. Finally, we explore the possibility of anomalous Hall effect due to this weak ferromagnetism. Our studies help thoroughly understand the electronic and magnetic structure of bulk orthorhombic BFO and CMO, shedding light on antiferromagnetic spin-splitting phenomena. Additionally, we illustrate the implications of spin-orbit interaction in these compounds in the forms of weak ferromagnetism and a nonvanishing anomalous Hall conductivity.

% ========== ACKNOWLEDGMENT ==========
\begin{acknowledgements}
S.R.\ and S.S.\ acknowledge research fellowships from CSIR, India, through grant number 09/1020(0157)/2019-EMR-I, and from UGC, India, through UGC-Ref. No. 1470, respectively. N.G.\ acknowledges the research fund from SERB, India, through grant number CRG/2021/005320. The use of high-performance computing facilities at IISER Bhopal and PARAM Seva within the framework of the National Supercomputing Mission, India, is gratefully acknowledged.
\end{acknowledgements}
% ====================

% ========== BIBLIOGRAPHY ==========
%\bibliography{library,references}

\begin{thebibliography}{51}%
\makeatletter
\providecommand \@ifxundefined [1]{%
 \@ifx{#1\undefined}
}%
\providecommand \@ifnum [1]{%
 \ifnum #1\expandafter \@firstoftwo
 \else \expandafter \@secondoftwo
 \fi
}%
\providecommand \@ifx [1]{%
 \ifx #1\expandafter \@firstoftwo
 \else \expandafter \@secondoftwo
 \fi
}%
\providecommand \natexlab [1]{#1}%
\providecommand \enquote  [1]{``#1''}%
\providecommand \bibnamefont  [1]{#1}%
\providecommand \bibfnamefont [1]{#1}%
\providecommand \citenamefont [1]{#1}%
\providecommand \href@noop [0]{\@secondoftwo}%
\providecommand \href [0]{\begingroup \@sanitize@url \@href}%
\providecommand \@href[1]{\@@startlink{#1}\@@href}%
\providecommand \@@href[1]{\endgroup#1\@@endlink}%
\providecommand \@sanitize@url [0]{\catcode `\\12\catcode `\$12\catcode
  `\&12\catcode `\#12\catcode `\^12\catcode `\_12\catcode `\%12\relax}%
\providecommand \@@startlink[1]{}%
\providecommand \@@endlink[0]{}%
\providecommand \url  [0]{\begingroup\@sanitize@url \@url }%
\providecommand \@url [1]{\endgroup\@href {#1}{\urlprefix }}%
\providecommand \urlprefix  [0]{URL }%
\providecommand \Eprint [0]{\href }%
\providecommand \doibase [0]{https://doi.org/}%
\providecommand \selectlanguage [0]{\@gobble}%
\providecommand \bibinfo  [0]{\@secondoftwo}%
\providecommand \bibfield  [0]{\@secondoftwo}%
\providecommand \translation [1]{[#1]}%
\providecommand \BibitemOpen [0]{}%
\providecommand \bibitemStop [0]{}%
\providecommand \bibitemNoStop [0]{.\EOS\space}%
\providecommand \EOS [0]{\spacefactor3000\relax}%
\providecommand \BibitemShut  [1]{\csname bibitem#1\endcsname}%
\let\auto@bib@innerbib\@empty
%</preamble>
\bibitem [{\citenamefont {Keimer}\ and\ \citenamefont
  {Moore}(2017)}]{KeimerNP17}%
  \BibitemOpen
  \bibfield  {author} {\bibinfo {author} {\bibfnamefont {B.}~\bibnamefont
  {Keimer}}\ and\ \bibinfo {author} {\bibfnamefont {J.~E.}\ \bibnamefont
  {Moore}},\ }\bibfield  {title} {\bibinfo {title} {{The physics of quantum
  materials}},\ }\href {https://doi.org/10.1038/nphys4302} {\bibfield
  {journal} {\bibinfo  {journal} {Nat. Phys.}\ }\textbf {\bibinfo {volume}
  {13}},\ \bibinfo {pages} {1045} (\bibinfo {year} {2017})}\BibitemShut
  {NoStop}%
\bibitem [{\citenamefont {Manchon}\ \emph {et~al.}(2015)\citenamefont
  {Manchon}, \citenamefont {Koo}, \citenamefont {Nitta}, \citenamefont
  {Frolov},\ and\ \citenamefont {Duine}}]{Manchon2015}%
  \BibitemOpen
  \bibfield  {author} {\bibinfo {author} {\bibfnamefont {A.}~\bibnamefont
  {Manchon}}, \bibinfo {author} {\bibfnamefont {H.~C.}\ \bibnamefont {Koo}},
  \bibinfo {author} {\bibfnamefont {J.}~\bibnamefont {Nitta}}, \bibinfo
  {author} {\bibfnamefont {S.~M.}\ \bibnamefont {Frolov}},\ and\ \bibinfo
  {author} {\bibfnamefont {R.~A.}\ \bibnamefont {Duine}},\ }\bibfield  {title}
  {\bibinfo {title} {{New perspectives for Rashba spin–orbit coupling}},\
  }\href {https://doi.org/10.1038/nmat4360} {\bibfield  {journal} {\bibinfo
  {journal} {Nat. Materials.}\ }\textbf {\bibinfo {volume} {14}},\ \bibinfo
  {pages} {1476} (\bibinfo {year} {2015})}\BibitemShut {NoStop}%
\bibitem [{\citenamefont {Baltz}\ \emph {et~al.}(2018)\citenamefont {Baltz},
  \citenamefont {Manchon}, \citenamefont {Tsoi}, \citenamefont {Moriyama},
  \citenamefont {Ono},\ and\ \citenamefont
  {Tserkovnyak}}]{RevModPhys.90.015005}%
  \BibitemOpen
  \bibfield  {author} {\bibinfo {author} {\bibfnamefont {V.}~\bibnamefont
  {Baltz}}, \bibinfo {author} {\bibfnamefont {A.}~\bibnamefont {Manchon}},
  \bibinfo {author} {\bibfnamefont {M.}~\bibnamefont {Tsoi}}, \bibinfo {author}
  {\bibfnamefont {T.}~\bibnamefont {Moriyama}}, \bibinfo {author}
  {\bibfnamefont {T.}~\bibnamefont {Ono}},\ and\ \bibinfo {author}
  {\bibfnamefont {Y.}~\bibnamefont {Tserkovnyak}},\ }\bibfield  {title}
  {\bibinfo {title} {Antiferromagnetic spintronics},\ }\href
  {https://doi.org/10.1103/RevModPhys.90.015005} {\bibfield  {journal}
  {\bibinfo  {journal} {Rev. Mod. Phys.}\ }\textbf {\bibinfo {volume} {90}},\
  \bibinfo {pages} {015005} (\bibinfo {year} {2018})}\BibitemShut {NoStop}%
\bibitem [{\citenamefont {Rooj}\ \emph {et~al.}(2024)\citenamefont {Rooj},
  \citenamefont {Chakraborty},\ and\ \citenamefont {Ganguli}}]{RoojAPR23}%
  \BibitemOpen
  \bibfield  {author} {\bibinfo {author} {\bibfnamefont {S.}~\bibnamefont
  {Rooj}}, \bibinfo {author} {\bibfnamefont {J.}~\bibnamefont {Chakraborty}},\
  and\ \bibinfo {author} {\bibfnamefont {N.}~\bibnamefont {Ganguli}},\
  }\bibfield  {title} {\bibinfo {title} {{Hexagonal MnTe with Antiferromagnetic
  Spin Splitting and Hidden Rashba - Dresselhaus Interaction for
  Antiferromagnetic Spintronics}},\ }\href
  {https://doi.org/10.1002/apxr.202300050} {\bibfield  {journal} {\bibinfo
  {journal} {Adv. Phys. Res.}\ }\textbf {\bibinfo {volume} {3}},\ \bibinfo
  {pages} {2300050} (\bibinfo {year} {2024})}\BibitemShut {NoStop}%
\bibitem [{\citenamefont {Šmejkal}\ \emph {et~al.}(2020)\citenamefont
  {Šmejkal}, \citenamefont {González-Hernández}, \citenamefont {Jungwirth},\
  and\ \citenamefont {Sinova}}]{Libor2020}%
  \BibitemOpen
  \bibfield  {author} {\bibinfo {author} {\bibfnamefont {L.}~\bibnamefont
  {Šmejkal}}, \bibinfo {author} {\bibfnamefont {R.}~\bibnamefont
  {González-Hernández}}, \bibinfo {author} {\bibfnamefont {T.}~\bibnamefont
  {Jungwirth}},\ and\ \bibinfo {author} {\bibfnamefont {J.}~\bibnamefont
  {Sinova}},\ }\bibfield  {title} {\bibinfo {title} {Crystal time-reversal
  symmetry breaking and spontaneous hall effect in collinear
  antiferromagnets},\ }\href {https://doi.org/10.1126/sciadv.aaz8809}
  {\bibfield  {journal} {\bibinfo  {journal} {Science Advances}\ }\textbf
  {\bibinfo {volume} {6}},\ \bibinfo {pages} {aaz8809} (\bibinfo {year}
  {2020})}\BibitemShut {NoStop}%
\bibitem [{\citenamefont {\ifmmode~\check{S}\else \v{S}\fi{}mejkal}\ \emph
  {et~al.}(2022{\natexlab{a}})\citenamefont {\ifmmode~\check{S}\else
  \v{S}\fi{}mejkal}, \citenamefont {Sinova},\ and\ \citenamefont
  {Jungwirth}}]{PhysRevX.12.031042}%
  \BibitemOpen
  \bibfield  {author} {\bibinfo {author} {\bibfnamefont {L.}~\bibnamefont
  {\ifmmode~\check{S}\else \v{S}\fi{}mejkal}}, \bibinfo {author} {\bibfnamefont
  {J.}~\bibnamefont {Sinova}},\ and\ \bibinfo {author} {\bibfnamefont
  {T.}~\bibnamefont {Jungwirth}},\ }\bibfield  {title} {\bibinfo {title}
  {Beyond conventional ferromagnetism and antiferromagnetism: A phase with
  nonrelativistic spin and crystal rotation symmetry},\ }\href
  {https://doi.org/10.1103/PhysRevX.12.031042} {\bibfield  {journal} {\bibinfo
  {journal} {Phys. Rev. X}\ }\textbf {\bibinfo {volume} {12}},\ \bibinfo
  {pages} {031042} (\bibinfo {year} {2022}{\natexlab{a}})}\BibitemShut
  {NoStop}%
\bibitem [{\citenamefont {\ifmmode~\check{S}\else \v{S}\fi{}mejkal}\ \emph
  {et~al.}(2022{\natexlab{b}})\citenamefont {\ifmmode~\check{S}\else
  \v{S}\fi{}mejkal}, \citenamefont {Sinova},\ and\ \citenamefont
  {Jungwirth}}]{PhysRevX.12.040501}%
  \BibitemOpen
  \bibfield  {author} {\bibinfo {author} {\bibfnamefont {L.}~\bibnamefont
  {\ifmmode~\check{S}\else \v{S}\fi{}mejkal}}, \bibinfo {author} {\bibfnamefont
  {J.}~\bibnamefont {Sinova}},\ and\ \bibinfo {author} {\bibfnamefont
  {T.}~\bibnamefont {Jungwirth}},\ }\bibfield  {title} {\bibinfo {title}
  {Emerging research landscape of altermagnetism},\ }\href
  {https://doi.org/10.1103/PhysRevX.12.040501} {\bibfield  {journal} {\bibinfo
  {journal} {Phys. Rev. X}\ }\textbf {\bibinfo {volume} {12}},\ \bibinfo
  {pages} {040501} (\bibinfo {year} {2022}{\natexlab{b}})}\BibitemShut
  {NoStop}%
\bibitem [{\citenamefont {Yuan}\ \emph {et~al.}(2021)\citenamefont {Yuan},
  \citenamefont {Wang}, \citenamefont {Luo},\ and\ \citenamefont
  {Zunger}}]{PhysRevMaterials.5.014409}%
  \BibitemOpen
  \bibfield  {author} {\bibinfo {author} {\bibfnamefont {L.~D.}\ \bibnamefont
  {Yuan}}, \bibinfo {author} {\bibfnamefont {Z.}~\bibnamefont {Wang}}, \bibinfo
  {author} {\bibfnamefont {J.~W.}\ \bibnamefont {Luo}},\ and\ \bibinfo {author}
  {\bibfnamefont {A.}~\bibnamefont {Zunger}},\ }\bibfield  {title} {\bibinfo
  {title} {{Prediction of low-Z collinear and noncollinear antiferromagnetic
  compounds having momentum-dependent spin splitting even without spin-orbit
  coupling}},\ }\href {https://doi.org/10.1103/PhysRevMaterials.5.014409}
  {\bibfield  {journal} {\bibinfo  {journal} {Physical Review Materials}\
  }\textbf {\bibinfo {volume} {5}},\ \bibinfo {pages} {14409} (\bibinfo {year}
  {2021})}\BibitemShut {NoStop}%
\bibitem [{\citenamefont {Yuan}\ \emph {et~al.}(2020)\citenamefont {Yuan},
  \citenamefont {Wang}, \citenamefont {Luo}, \citenamefont {Rashba},\ and\
  \citenamefont {Zunger}}]{PhysRevB.102.014422}%
  \BibitemOpen
  \bibfield  {author} {\bibinfo {author} {\bibfnamefont {L.~D.}\ \bibnamefont
  {Yuan}}, \bibinfo {author} {\bibfnamefont {Z.}~\bibnamefont {Wang}}, \bibinfo
  {author} {\bibfnamefont {J.~W.}\ \bibnamefont {Luo}}, \bibinfo {author}
  {\bibfnamefont {E.~I.}\ \bibnamefont {Rashba}},\ and\ \bibinfo {author}
  {\bibfnamefont {A.}~\bibnamefont {Zunger}},\ }\bibfield  {title} {\bibinfo
  {title} {{Giant momentum-dependent spin splitting in centrosymmetric low- Z
  antiferromagnets}},\ }\href {https://doi.org/10.1103/PhysRevB.102.014422}
  {\bibfield  {journal} {\bibinfo  {journal} {Physical Review B}\ }\textbf
  {\bibinfo {volume} {102}},\ \bibinfo {pages} {14422} (\bibinfo {year}
  {2020})}\BibitemShut {NoStop}%
\bibitem [{\citenamefont {Hayami}\ \emph {et~al.}(2020)\citenamefont {Hayami},
  \citenamefont {Yanagi},\ and\ \citenamefont {Kusunose}}]{Hayami2020}%
  \BibitemOpen
  \bibfield  {author} {\bibinfo {author} {\bibfnamefont {S.}~\bibnamefont
  {Hayami}}, \bibinfo {author} {\bibfnamefont {Y.}~\bibnamefont {Yanagi}},\
  and\ \bibinfo {author} {\bibfnamefont {H.}~\bibnamefont {Kusunose}},\
  }\bibfield  {title} {\bibinfo {title} {{Bottom-up design of spin-split and
  reshaped electronic band structures in antiferromagnets without spin-orbit
  coupling: Procedure on the basis of augmented multipoles}},\ }\href
  {https://doi.org/10.1103/PhysRevB.102.144441} {\bibfield  {journal} {\bibinfo
   {journal} {Physical Review B}\ }\textbf {\bibinfo {volume} {102}},\ \bibinfo
  {pages} {144441} (\bibinfo {year} {2020})}\BibitemShut {NoStop}%
\bibitem [{\citenamefont {Gonz{\'{a}}lez-Hern{\'{a}}ndez}\ \emph
  {et~al.}(2021)\citenamefont {Gonz{\'{a}}lez-Hern{\'{a}}ndez}, \citenamefont
  {{\v{S}}mejkal}, \citenamefont {V{\'{y}}born{\'{y}}}, \citenamefont {Yahagi},
  \citenamefont {Sinova}, \citenamefont {Jungwirth},\ and\ \citenamefont
  {{\v{Z}}elezn{\'{y}}}}]{Hernandez2021}%
  \BibitemOpen
  \bibfield  {author} {\bibinfo {author} {\bibfnamefont {R.}~\bibnamefont
  {Gonz{\'{a}}lez-Hern{\'{a}}ndez}}, \bibinfo {author} {\bibfnamefont
  {L.}~\bibnamefont {{\v{S}}mejkal}}, \bibinfo {author} {\bibfnamefont
  {K.}~\bibnamefont {V{\'{y}}born{\'{y}}}}, \bibinfo {author} {\bibfnamefont
  {Y.}~\bibnamefont {Yahagi}}, \bibinfo {author} {\bibfnamefont
  {J.}~\bibnamefont {Sinova}}, \bibinfo {author} {\bibfnamefont
  {T.}~\bibnamefont {Jungwirth}},\ and\ \bibinfo {author} {\bibfnamefont
  {J.}~\bibnamefont {{\v{Z}}elezn{\'{y}}}},\ }\bibfield  {title} {\bibinfo
  {title} {{Efficient Electrical Spin Splitter Based on Nonrelativistic
  Collinear Antiferromagnetism}},\ }\href
  {https://doi.org/10.1103/PhysRevLett.126.127701} {\bibfield  {journal}
  {\bibinfo  {journal} {Physical Review Letters}\ }\textbf {\bibinfo {volume}
  {126}},\ \bibinfo {pages} {127701} (\bibinfo {year} {2021})}\BibitemShut
  {NoStop}%
\bibitem [{\citenamefont {Bai}\ \emph {et~al.}(2022)\citenamefont {Bai},
  \citenamefont {Han}, \citenamefont {Feng}, \citenamefont {Zhou},
  \citenamefont {Su}, \citenamefont {Wang}, \citenamefont {Liao}, \citenamefont
  {Zhu}, \citenamefont {Chen}, \citenamefont {Pan}, \citenamefont {Fan},\ and\
  \citenamefont {Song}}]{HBai2022}%
  \BibitemOpen
  \bibfield  {author} {\bibinfo {author} {\bibfnamefont {H.}~\bibnamefont
  {Bai}}, \bibinfo {author} {\bibfnamefont {L.}~\bibnamefont {Han}}, \bibinfo
  {author} {\bibfnamefont {X.~Y.}\ \bibnamefont {Feng}}, \bibinfo {author}
  {\bibfnamefont {Y.~J.}\ \bibnamefont {Zhou}}, \bibinfo {author}
  {\bibfnamefont {R.~X.}\ \bibnamefont {Su}}, \bibinfo {author} {\bibfnamefont
  {Q.}~\bibnamefont {Wang}}, \bibinfo {author} {\bibfnamefont {L.~Y.}\
  \bibnamefont {Liao}}, \bibinfo {author} {\bibfnamefont {W.~X.}\ \bibnamefont
  {Zhu}}, \bibinfo {author} {\bibfnamefont {X.~Z.}\ \bibnamefont {Chen}},
  \bibinfo {author} {\bibfnamefont {F.}~\bibnamefont {Pan}}, \bibinfo {author}
  {\bibfnamefont {X.~L.}\ \bibnamefont {Fan}},\ and\ \bibinfo {author}
  {\bibfnamefont {C.}~\bibnamefont {Song}},\ }\bibfield  {title} {\bibinfo
  {title} {Observation of spin splitting torque in a collinear antiferromagnet
  {RuO$_2$}},\ }\href {https://doi.org/10.1103/PhysRevLett.128.197202}
  {\bibfield  {journal} {\bibinfo  {journal} {Phys. Rev. Lett.}\ }\textbf
  {\bibinfo {volume} {128}},\ \bibinfo {pages} {197202} (\bibinfo {year}
  {2022})}\BibitemShut {NoStop}%
\bibitem [{\citenamefont {Ke{\ss}ler}\ \emph {et~al.}(2024)\citenamefont
  {Ke{\ss}ler}, \citenamefont {Garcia-Gassull}, \citenamefont {Suter},
  \citenamefont {Prokscha}, \citenamefont {Salman}, \citenamefont {Khalyavin},
  \citenamefont {Manuel}, \citenamefont {Orlandi}, \citenamefont {Mazin},
  \citenamefont {Valent{\'{i}}},\ and\ \citenamefont {Moser}}]{KesslerNPJS24}%
  \BibitemOpen
  \bibfield  {author} {\bibinfo {author} {\bibfnamefont {P.}~\bibnamefont
  {Ke{\ss}ler}}, \bibinfo {author} {\bibfnamefont {L.}~\bibnamefont
  {Garcia-Gassull}}, \bibinfo {author} {\bibfnamefont {A.}~\bibnamefont
  {Suter}}, \bibinfo {author} {\bibfnamefont {T.}~\bibnamefont {Prokscha}},
  \bibinfo {author} {\bibfnamefont {Z.}~\bibnamefont {Salman}}, \bibinfo
  {author} {\bibfnamefont {D.}~\bibnamefont {Khalyavin}}, \bibinfo {author}
  {\bibfnamefont {P.}~\bibnamefont {Manuel}}, \bibinfo {author} {\bibfnamefont
  {F.}~\bibnamefont {Orlandi}}, \bibinfo {author} {\bibfnamefont {I.~I.}\
  \bibnamefont {Mazin}}, \bibinfo {author} {\bibfnamefont {R.}~\bibnamefont
  {Valent{\'{i}}}},\ and\ \bibinfo {author} {\bibfnamefont {S.}~\bibnamefont
  {Moser}},\ }\bibfield  {title} {\bibinfo {title} {{Absence of magnetic order
  in RuO$_2$: insights from $\mu$SR spectroscopy and neutron diffraction}},\
  }\href {https://doi.org/10.1038/s44306-024-00055-y} {\bibfield  {journal}
  {\bibinfo  {journal} {npj Spintronics}\ }\textbf {\bibinfo {volume} {2}},\
  \bibinfo {pages} {50} (\bibinfo {year} {2024})}\BibitemShut {NoStop}%
\bibitem [{\citenamefont {Liu}\ \emph {et~al.}(2024)\citenamefont {Liu},
  \citenamefont {Zhan}, \citenamefont {Li}, \citenamefont {Liu}, \citenamefont
  {Cheng}, \citenamefont {Shi}, \citenamefont {Deng}, \citenamefont {Zhang},
  \citenamefont {Li}, \citenamefont {Ding}, \citenamefont {Jiang},
  \citenamefont {Ye}, \citenamefont {Liu}, \citenamefont {Jiang}, \citenamefont
  {Wang}, \citenamefont {Li}, \citenamefont {Xie}, \citenamefont {Wang},
  \citenamefont {Qiao}, \citenamefont {Wen}, \citenamefont {Sun},\ and\
  \citenamefont {Shen}}]{LiuPRL24}%
  \BibitemOpen
  \bibfield  {author} {\bibinfo {author} {\bibfnamefont {J.}~\bibnamefont
  {Liu}}, \bibinfo {author} {\bibfnamefont {J.}~\bibnamefont {Zhan}}, \bibinfo
  {author} {\bibfnamefont {T.}~\bibnamefont {Li}}, \bibinfo {author}
  {\bibfnamefont {J.}~\bibnamefont {Liu}}, \bibinfo {author} {\bibfnamefont
  {S.}~\bibnamefont {Cheng}}, \bibinfo {author} {\bibfnamefont
  {Y.}~\bibnamefont {Shi}}, \bibinfo {author} {\bibfnamefont {L.}~\bibnamefont
  {Deng}}, \bibinfo {author} {\bibfnamefont {M.}~\bibnamefont {Zhang}},
  \bibinfo {author} {\bibfnamefont {C.}~\bibnamefont {Li}}, \bibinfo {author}
  {\bibfnamefont {J.}~\bibnamefont {Ding}}, \bibinfo {author} {\bibfnamefont
  {Q.}~\bibnamefont {Jiang}}, \bibinfo {author} {\bibfnamefont
  {M.}~\bibnamefont {Ye}}, \bibinfo {author} {\bibfnamefont {Z.}~\bibnamefont
  {Liu}}, \bibinfo {author} {\bibfnamefont {Z.}~\bibnamefont {Jiang}}, \bibinfo
  {author} {\bibfnamefont {S.}~\bibnamefont {Wang}}, \bibinfo {author}
  {\bibfnamefont {Q.}~\bibnamefont {Li}}, \bibinfo {author} {\bibfnamefont
  {Y.}~\bibnamefont {Xie}}, \bibinfo {author} {\bibfnamefont {Y.}~\bibnamefont
  {Wang}}, \bibinfo {author} {\bibfnamefont {S.}~\bibnamefont {Qiao}}, \bibinfo
  {author} {\bibfnamefont {J.}~\bibnamefont {Wen}}, \bibinfo {author}
  {\bibfnamefont {Y.}~\bibnamefont {Sun}},\ and\ \bibinfo {author}
  {\bibfnamefont {D.}~\bibnamefont {Shen}},\ }\bibfield  {title} {\bibinfo
  {title} {{Absence of Altermagnetic Spin Splitting Character in Rutile Oxide
  RuO$_2$}},\ }\href {https://doi.org/10.1103/PhysRevLett.133.176401}
  {\bibfield  {journal} {\bibinfo  {journal} {Phys. Rev. Lett.}\ }\textbf
  {\bibinfo {volume} {133}},\ \bibinfo {pages} {176401} (\bibinfo {year}
  {2024})}\BibitemShut {NoStop}%
\bibitem [{\citenamefont {{Gonzalez Betancourt}}\ \emph
  {et~al.}(2023)\citenamefont {{Gonzalez Betancourt}}, \citenamefont
  {Zub{\'{a}}{\v{c}}}, \citenamefont {Gonzalez-Hernandez}, \citenamefont
  {Geishendorf}, \citenamefont {{\v{S}}ob{\'{a}}ň}, \citenamefont
  {Springholz}, \citenamefont {Olejn{\'{i}}k}, \citenamefont {{\v{S}}mejkal},
  \citenamefont {Sinova}, \citenamefont {Jungwirth}, \citenamefont
  {Goennenwein}, \citenamefont {Thomas}, \citenamefont {Reichlov{\'{a}}},
  \citenamefont {{\v{Z}}elezn{\'{y}}},\ and\ \citenamefont
  {Kriegner}}]{MnTeAHE2023}%
  \BibitemOpen
  \bibfield  {author} {\bibinfo {author} {\bibfnamefont {R.~D.}\ \bibnamefont
  {{Gonzalez Betancourt}}}, \bibinfo {author} {\bibfnamefont {J.}~\bibnamefont
  {Zub{\'{a}}{\v{c}}}}, \bibinfo {author} {\bibfnamefont {R.}~\bibnamefont
  {Gonzalez-Hernandez}}, \bibinfo {author} {\bibfnamefont {K.}~\bibnamefont
  {Geishendorf}}, \bibinfo {author} {\bibfnamefont {Z.}~\bibnamefont
  {{\v{S}}ob{\'{a}}ň}}, \bibinfo {author} {\bibfnamefont {G.}~\bibnamefont
  {Springholz}}, \bibinfo {author} {\bibfnamefont {K.}~\bibnamefont
  {Olejn{\'{i}}k}}, \bibinfo {author} {\bibfnamefont {L.}~\bibnamefont
  {{\v{S}}mejkal}}, \bibinfo {author} {\bibfnamefont {J.}~\bibnamefont
  {Sinova}}, \bibinfo {author} {\bibfnamefont {T.}~\bibnamefont {Jungwirth}},
  \bibinfo {author} {\bibfnamefont {S.~T.}\ \bibnamefont {Goennenwein}},
  \bibinfo {author} {\bibfnamefont {A.}~\bibnamefont {Thomas}}, \bibinfo
  {author} {\bibfnamefont {H.}~\bibnamefont {Reichlov{\'{a}}}}, \bibinfo
  {author} {\bibfnamefont {J.}~\bibnamefont {{\v{Z}}elezn{\'{y}}}},\ and\
  \bibinfo {author} {\bibfnamefont {D.}~\bibnamefont {Kriegner}},\ }\bibfield
  {title} {\bibinfo {title} {{Spontaneous Anomalous Hall Effect Arising from an
  Unconventional Compensated Magnetic Phase in a Semiconductor}},\ }\href
  {https://doi.org/10.1103/PhysRevLett.130.036702} {\bibfield  {journal}
  {\bibinfo  {journal} {Physical Review Letters}\ }\textbf {\bibinfo {volume}
  {130}},\ \bibinfo {pages} {036702} (\bibinfo {year} {2023})}\BibitemShut
  {NoStop}%
\bibitem [{\citenamefont {Feng}\ \emph {et~al.}(2022)\citenamefont {Feng},
  \citenamefont {Zhou}, \citenamefont {Šmejkal}, \citenamefont {Wu},
  \citenamefont {Zhu}, \citenamefont {Guo}, \citenamefont
  {González-Hernández}, \citenamefont {Wang}, \citenamefont {Yan},
  \citenamefont {Qin}, \citenamefont {Zhang}, \citenamefont {Wu}, \citenamefont
  {Chen}, \citenamefont {Meng}, \citenamefont {Liu}, \citenamefont {Xia},
  \citenamefont {Sinova}, \citenamefont {Jungwirth},\ and\ \citenamefont
  {Liu}}]{Feng2022}%
  \BibitemOpen
  \bibfield  {author} {\bibinfo {author} {\bibfnamefont {Z.}~\bibnamefont
  {Feng}}, \bibinfo {author} {\bibfnamefont {X.}~\bibnamefont {Zhou}}, \bibinfo
  {author} {\bibfnamefont {L.}~\bibnamefont {Šmejkal}}, \bibinfo {author}
  {\bibfnamefont {L.}~\bibnamefont {Wu}}, \bibinfo {author} {\bibfnamefont
  {Z.}~\bibnamefont {Zhu}}, \bibinfo {author} {\bibfnamefont {H.}~\bibnamefont
  {Guo}}, \bibinfo {author} {\bibfnamefont {R.}~\bibnamefont
  {González-Hernández}}, \bibinfo {author} {\bibfnamefont {X.}~\bibnamefont
  {Wang}}, \bibinfo {author} {\bibfnamefont {H.}~\bibnamefont {Yan}}, \bibinfo
  {author} {\bibfnamefont {P.}~\bibnamefont {Qin}}, \bibinfo {author}
  {\bibfnamefont {X.}~\bibnamefont {Zhang}}, \bibinfo {author} {\bibfnamefont
  {H.}~\bibnamefont {Wu}}, \bibinfo {author} {\bibfnamefont {H.}~\bibnamefont
  {Chen}}, \bibinfo {author} {\bibfnamefont {Z.}~\bibnamefont {Meng}}, \bibinfo
  {author} {\bibfnamefont {L.}~\bibnamefont {Liu}}, \bibinfo {author}
  {\bibfnamefont {Z.}~\bibnamefont {Xia}}, \bibinfo {author} {\bibfnamefont
  {J.}~\bibnamefont {Sinova}}, \bibinfo {author} {\bibfnamefont
  {T.}~\bibnamefont {Jungwirth}},\ and\ \bibinfo {author} {\bibfnamefont
  {Z.}~\bibnamefont {Liu}},\ }\bibfield  {title} {\bibinfo {title} {{An
  anomalous Hall effect in altermagnetic ruthenium dioxide}},\ }\href
  {https://doi.org/10.1038/s41928-022-00866-z} {\bibfield  {journal} {\bibinfo
  {journal} {Nature Electronics}\ }\textbf {\bibinfo {volume} {5}},\ \bibinfo
  {pages} {2520} (\bibinfo {year} {2022})}\BibitemShut {NoStop}%
\bibitem [{\citenamefont {Kluczyk}\ \emph {et~al.}()\citenamefont {Kluczyk},
  \citenamefont {Gas}, \citenamefont {Grzybowski}, \citenamefont {Skupiński},
  \citenamefont {Borysiewicz}, \citenamefont {Fas}, \citenamefont
  {Suffczyński}, \citenamefont {Domagala}, \citenamefont {Grasza},
  \citenamefont {Mycielski}, \citenamefont {Baj}, \citenamefont {Ahn},
  \citenamefont {Výborný}, \citenamefont {Sawicki},\ and\ \citenamefont
  {Gryglas-Borysiewicz}}]{kluczyk2023}%
  \BibitemOpen
  \bibfield  {author} {\bibinfo {author} {\bibfnamefont {K.~P.}\ \bibnamefont
  {Kluczyk}}, \bibinfo {author} {\bibfnamefont {K.}~\bibnamefont {Gas}},
  \bibinfo {author} {\bibfnamefont {M.~J.}\ \bibnamefont {Grzybowski}},
  \bibinfo {author} {\bibfnamefont {P.}~\bibnamefont {Skupiński}}, \bibinfo
  {author} {\bibfnamefont {M.~A.}\ \bibnamefont {Borysiewicz}}, \bibinfo
  {author} {\bibfnamefont {T.}~\bibnamefont {Fas}}, \bibinfo {author}
  {\bibfnamefont {J.}~\bibnamefont {Suffczyński}}, \bibinfo {author}
  {\bibfnamefont {J.~Z.}\ \bibnamefont {Domagala}}, \bibinfo {author}
  {\bibfnamefont {K.}~\bibnamefont {Grasza}}, \bibinfo {author} {\bibfnamefont
  {A.}~\bibnamefont {Mycielski}}, \bibinfo {author} {\bibfnamefont
  {M.}~\bibnamefont {Baj}}, \bibinfo {author} {\bibfnamefont {K.~H.}\
  \bibnamefont {Ahn}}, \bibinfo {author} {\bibfnamefont {K.}~\bibnamefont
  {Výborný}}, \bibinfo {author} {\bibfnamefont {M.}~\bibnamefont {Sawicki}},\
  and\ \bibinfo {author} {\bibfnamefont {M.}~\bibnamefont
  {Gryglas-Borysiewicz}},\ }\bibfield  {title} {\bibinfo {title} {Coexistence
  of anomalous hall effect and weak net magnetization in collinear
  antiferromagnet {MnTe}},\ }\href@noop {} {\ }\Eprint
  {https://arxiv.org/abs/2310.09134 {(2023)}} {arXiv:2310.09134 {(2023)}}
  \BibitemShut {NoStop}%
\bibitem [{\citenamefont {Leiviskä}\ \emph {et~al.}()\citenamefont
  {Leiviskä}, \citenamefont {Rial}, \citenamefont {Badura}, \citenamefont
  {Seeger}, \citenamefont {Kounta}, \citenamefont {Beckert}, \citenamefont
  {Kriegner}, \citenamefont {Joumard}, \citenamefont {Schmoranzerová},
  \citenamefont {Sinova}, \citenamefont {Gomonay}, \citenamefont {Thomas},
  \citenamefont {Goennenwein}, \citenamefont {Reichlová}, \citenamefont
  {Šmejkal}, \citenamefont {Michez}, \citenamefont {Jungwirth},\ and\
  \citenamefont {Baltz}}]{leiviskä2024}%
  \BibitemOpen
  \bibfield  {author} {\bibinfo {author} {\bibfnamefont {M.}~\bibnamefont
  {Leiviskä}}, \bibinfo {author} {\bibfnamefont {J.}~\bibnamefont {Rial}},
  \bibinfo {author} {\bibfnamefont {A.}~\bibnamefont {Badura}}, \bibinfo
  {author} {\bibfnamefont {R.~L.}\ \bibnamefont {Seeger}}, \bibinfo {author}
  {\bibfnamefont {I.}~\bibnamefont {Kounta}}, \bibinfo {author} {\bibfnamefont
  {S.}~\bibnamefont {Beckert}}, \bibinfo {author} {\bibfnamefont
  {D.}~\bibnamefont {Kriegner}}, \bibinfo {author} {\bibfnamefont
  {I.}~\bibnamefont {Joumard}}, \bibinfo {author} {\bibfnamefont
  {E.}~\bibnamefont {Schmoranzerová}}, \bibinfo {author} {\bibfnamefont
  {J.}~\bibnamefont {Sinova}}, \bibinfo {author} {\bibfnamefont
  {O.}~\bibnamefont {Gomonay}}, \bibinfo {author} {\bibfnamefont
  {A.}~\bibnamefont {Thomas}}, \bibinfo {author} {\bibfnamefont {S.~T.~B.}\
  \bibnamefont {Goennenwein}}, \bibinfo {author} {\bibfnamefont
  {H.}~\bibnamefont {Reichlová}}, \bibinfo {author} {\bibfnamefont
  {L.}~\bibnamefont {Šmejkal}}, \bibinfo {author} {\bibfnamefont
  {L.}~\bibnamefont {Michez}}, \bibinfo {author} {\bibfnamefont
  {T.}~\bibnamefont {Jungwirth}},\ and\ \bibinfo {author} {\bibfnamefont
  {V.}~\bibnamefont {Baltz}},\ }\bibfield  {title} {\bibinfo {title}
  {Anisotropy of the anomalous hall effect in the altermagnet candidate
  {Mn$_5$Si$_3$} films},\ }\href@noop {} {\ }\Eprint
  {https://arxiv.org/abs/2401.02275 {(2024)}} {arXiv:2401.02275 {(2024)}}
  \BibitemShut {NoStop}%
\bibitem [{\citenamefont {Jiang}\ \emph {et~al.}(2023)\citenamefont {Jiang},
  \citenamefont {Wang}, \citenamefont {Samanta}, \citenamefont {Zhang},
  \citenamefont {Xiao}, \citenamefont {Lu}, \citenamefont {Sun}, \citenamefont
  {Tsymbal},\ and\ \citenamefont {Shao}}]{Jiang2023}%
  \BibitemOpen
  \bibfield  {author} {\bibinfo {author} {\bibfnamefont {Y.~Y.}\ \bibnamefont
  {Jiang}}, \bibinfo {author} {\bibfnamefont {Z.~A.}\ \bibnamefont {Wang}},
  \bibinfo {author} {\bibfnamefont {K.}~\bibnamefont {Samanta}}, \bibinfo
  {author} {\bibfnamefont {S.~H.}\ \bibnamefont {Zhang}}, \bibinfo {author}
  {\bibfnamefont {R.~C.}\ \bibnamefont {Xiao}}, \bibinfo {author}
  {\bibfnamefont {W.~J.}\ \bibnamefont {Lu}}, \bibinfo {author} {\bibfnamefont
  {Y.~P.}\ \bibnamefont {Sun}}, \bibinfo {author} {\bibfnamefont {E.~Y.}\
  \bibnamefont {Tsymbal}},\ and\ \bibinfo {author} {\bibfnamefont {D.~F.}\
  \bibnamefont {Shao}},\ }\bibfield  {title} {\bibinfo {title} {{Prediction of
  giant tunneling magnetoresistance in {RuO$_{2}$$|$TiO$_{2}$$|$RuO$_2$} (110)
  antiferromagnetic tunnel junctions}},\ }\href
  {https://doi.org/10.1103/PhysRevB.108.174439} {\bibfield  {journal} {\bibinfo
   {journal} {Physical Review B}\ }\textbf {\bibinfo {volume} {108}},\ \bibinfo
  {pages} {174439} (\bibinfo {year} {2023})}\BibitemShut {NoStop}%
\bibitem [{\citenamefont {{\v{S}}mejkal}\ \emph {et~al.}(2023)\citenamefont
  {{\v{S}}mejkal}, \citenamefont {Marmodoro}, \citenamefont {Ahn},
  \citenamefont {Gonz{\'{a}}lez-Hern{\'{a}}ndez}, \citenamefont {Turek},
  \citenamefont {Mankovsky}, \citenamefont {Ebert}, \citenamefont {D'Souza},
  \citenamefont {{\v{S}}ipr}, \citenamefont {Sinova},\ and\ \citenamefont
  {Jungwirth}}]{LiborMagnon2023}%
  \BibitemOpen
  \bibfield  {author} {\bibinfo {author} {\bibfnamefont {L.}~\bibnamefont
  {{\v{S}}mejkal}}, \bibinfo {author} {\bibfnamefont {A.}~\bibnamefont
  {Marmodoro}}, \bibinfo {author} {\bibfnamefont {K.~H.}\ \bibnamefont {Ahn}},
  \bibinfo {author} {\bibfnamefont {R.}~\bibnamefont
  {Gonz{\'{a}}lez-Hern{\'{a}}ndez}}, \bibinfo {author} {\bibfnamefont
  {I.}~\bibnamefont {Turek}}, \bibinfo {author} {\bibfnamefont
  {S.}~\bibnamefont {Mankovsky}}, \bibinfo {author} {\bibfnamefont
  {H.}~\bibnamefont {Ebert}}, \bibinfo {author} {\bibfnamefont {S.~W.}\
  \bibnamefont {D'Souza}}, \bibinfo {author} {\bibfnamefont {O.}~\bibnamefont
  {{\v{S}}ipr}}, \bibinfo {author} {\bibfnamefont {J.}~\bibnamefont {Sinova}},\
  and\ \bibinfo {author} {\bibfnamefont {T.}~\bibnamefont {Jungwirth}},\
  }\bibfield  {title} {\bibinfo {title} {Chiral magnons in altermagnetic
  {RuO$_2$}},\ }\href {https://doi.org/10.1103/PhysRevLett.131.256703}
  {\bibfield  {journal} {\bibinfo  {journal} {Physical Review Letters}\
  }\textbf {\bibinfo {volume} {131}},\ \bibinfo {pages} {256703} (\bibinfo
  {year} {2023})}\BibitemShut {NoStop}%
\bibitem [{\citenamefont {Zhou}\ \emph {et~al.}(2024)\citenamefont {Zhou},
  \citenamefont {Feng}, \citenamefont {Zhang}, \citenamefont {{\v{S}}mejkal},
  \citenamefont {Sinova}, \citenamefont {Mokrousov},\ and\ \citenamefont
  {Yao}}]{Libor2024}%
  \BibitemOpen
  \bibfield  {author} {\bibinfo {author} {\bibfnamefont {X.}~\bibnamefont
  {Zhou}}, \bibinfo {author} {\bibfnamefont {W.}~\bibnamefont {Feng}}, \bibinfo
  {author} {\bibfnamefont {R.~W.}\ \bibnamefont {Zhang}}, \bibinfo {author}
  {\bibfnamefont {L.}~\bibnamefont {{\v{S}}mejkal}}, \bibinfo {author}
  {\bibfnamefont {J.}~\bibnamefont {Sinova}}, \bibinfo {author} {\bibfnamefont
  {Y.}~\bibnamefont {Mokrousov}},\ and\ \bibinfo {author} {\bibfnamefont
  {Y.}~\bibnamefont {Yao}},\ }\bibfield  {title} {\bibinfo {title} {Crystal
  thermal transport in altermagnetic {RuO$_2$}},\ }\href
  {https://doi.org/10.1103/PhysRevLett.132.056701} {\bibfield  {journal}
  {\bibinfo  {journal} {Physical Review Letters}\ }\textbf {\bibinfo {volume}
  {132}},\ \bibinfo {pages} {056701} (\bibinfo {year} {2024})}\BibitemShut
  {NoStop}%
\bibitem [{\citenamefont {Osumi}\ \emph {et~al.}(2024)\citenamefont {Osumi},
  \citenamefont {Souma}, \citenamefont {Aoyama}, \citenamefont {Yamauchi},
  \citenamefont {Honma}, \citenamefont {Nakayama}, \citenamefont {Takahashi},
  \citenamefont {Ohgushi},\ and\ \citenamefont {Sato}}]{PhysRevB.109.115102}%
  \BibitemOpen
  \bibfield  {author} {\bibinfo {author} {\bibfnamefont {T.}~\bibnamefont
  {Osumi}}, \bibinfo {author} {\bibfnamefont {S.}~\bibnamefont {Souma}},
  \bibinfo {author} {\bibfnamefont {T.}~\bibnamefont {Aoyama}}, \bibinfo
  {author} {\bibfnamefont {K.}~\bibnamefont {Yamauchi}}, \bibinfo {author}
  {\bibfnamefont {A.}~\bibnamefont {Honma}}, \bibinfo {author} {\bibfnamefont
  {K.}~\bibnamefont {Nakayama}}, \bibinfo {author} {\bibfnamefont
  {T.}~\bibnamefont {Takahashi}}, \bibinfo {author} {\bibfnamefont
  {K.}~\bibnamefont {Ohgushi}},\ and\ \bibinfo {author} {\bibfnamefont
  {T.}~\bibnamefont {Sato}},\ }\bibfield  {title} {\bibinfo {title}
  {Observation of a giant band splitting in altermagnetic {MnTe}},\ }\href
  {https://doi.org/10.1103/PhysRevB.109.115102} {\bibfield  {journal} {\bibinfo
   {journal} {Phys. Rev. B}\ }\textbf {\bibinfo {volume} {109}},\ \bibinfo
  {pages} {115102} (\bibinfo {year} {2024})}\BibitemShut {NoStop}%
\bibitem [{\citenamefont {Milivojevi{\'{c}}}\ \emph {et~al.}(2024)\citenamefont
  {Milivojevi{\'{c}}}, \citenamefont {Orozovi{\'{c}}}, \citenamefont {Picozzi},
  \citenamefont {Gmitra},\ and\ \citenamefont
  {Stavri{\'{c}}}}]{Milivojevic2DM24}%
  \BibitemOpen
  \bibfield  {author} {\bibinfo {author} {\bibfnamefont {M.}~\bibnamefont
  {Milivojevi{\'{c}}}}, \bibinfo {author} {\bibfnamefont {M.}~\bibnamefont
  {Orozovi{\'{c}}}}, \bibinfo {author} {\bibfnamefont {S.}~\bibnamefont
  {Picozzi}}, \bibinfo {author} {\bibfnamefont {M.}~\bibnamefont {Gmitra}},\
  and\ \bibinfo {author} {\bibfnamefont {S.}~\bibnamefont {Stavri{\'{c}}}},\
  }\bibfield  {title} {\bibinfo {title} {{Interplay of altermagnetism and weak
  ferromagnetism in two-dimensional RuF$_4$}},\ }\href
  {https://doi.org/10.1088/2053-1583/ad4c73} {\bibfield  {journal} {\bibinfo
  {journal} {2D Mater.}\ }\textbf {\bibinfo {volume} {11}},\ \bibinfo {pages}
  {035025} (\bibinfo {year} {2024})}\BibitemShut {NoStop}%
\bibitem [{\citenamefont {Fakhredine}\ \emph {et~al.}(2023)\citenamefont
  {Fakhredine}, \citenamefont {Sattigeri}, \citenamefont {Cuono},\ and\
  \citenamefont {Autieri}}]{FakhredinePRB23}%
  \BibitemOpen
  \bibfield  {author} {\bibinfo {author} {\bibfnamefont {A.}~\bibnamefont
  {Fakhredine}}, \bibinfo {author} {\bibfnamefont {R.~M.}\ \bibnamefont
  {Sattigeri}}, \bibinfo {author} {\bibfnamefont {G.}~\bibnamefont {Cuono}},\
  and\ \bibinfo {author} {\bibfnamefont {C.}~\bibnamefont {Autieri}},\
  }\bibfield  {title} {\bibinfo {title} {{Interplay between altermagnetism and
  nonsymmorphic symmetries generating large anomalous Hall conductivity by
  semi-Dirac points induced anticrossings}},\ }\href
  {https://doi.org/10.1103/PhysRevB.108.115138} {\bibfield  {journal} {\bibinfo
   {journal} {Phys. Rev. B}\ }\textbf {\bibinfo {volume} {108}},\ \bibinfo
  {pages} {115138} (\bibinfo {year} {2023})}\BibitemShut {NoStop}%
\bibitem [{\citenamefont {Singh}\ \emph {et~al.}(2024)\citenamefont {Singh},
  \citenamefont {Cheong},\ and\ \citenamefont {Guo}}]{SinghARXIV24}%
  \BibitemOpen
  \bibfield  {author} {\bibinfo {author} {\bibfnamefont {D.~K.}\ \bibnamefont
  {Singh}}, \bibinfo {author} {\bibfnamefont {S.-W.}\ \bibnamefont {Cheong}},\
  and\ \bibinfo {author} {\bibfnamefont {J.}~\bibnamefont {Guo}},\ }\bibfield
  {title} {\bibinfo {title} {{Altermagnetism in NiSi and antiferromagnetic
  candidate materials with non-collinear spins}},\ }\href
  {https://doi.org/10.48550/arXiv.2402.17451} {\bibfield  {journal} {\bibinfo
  {journal} {arXiv}\ ,\ \bibinfo {pages} {2402.17451}} (\bibinfo {year}
  {2024})},\ \Eprint {https://arxiv.org/abs/2402.17451} {arXiv:2402.17451}
  \BibitemShut {NoStop}%
\bibitem [{\citenamefont {Kornev}\ \emph {et~al.}(2007)\citenamefont {Kornev},
  \citenamefont {Lisenkov}, \citenamefont {Haumont}, \citenamefont {Dkhil},\
  and\ \citenamefont {Bellaiche}}]{bfo_lattice}%
  \BibitemOpen
  \bibfield  {author} {\bibinfo {author} {\bibfnamefont {I.~A.}\ \bibnamefont
  {Kornev}}, \bibinfo {author} {\bibfnamefont {S.}~\bibnamefont {Lisenkov}},
  \bibinfo {author} {\bibfnamefont {R.}~\bibnamefont {Haumont}}, \bibinfo
  {author} {\bibfnamefont {B.}~\bibnamefont {Dkhil}},\ and\ \bibinfo {author}
  {\bibfnamefont {L.}~\bibnamefont {Bellaiche}},\ }\bibfield  {title} {\bibinfo
  {title} {Finite-temperature properties of multiferroic {BiFeO$_3$}},\ }\href
  {https://doi.org/10.1103/PhysRevLett.99.227602} {\bibfield  {journal}
  {\bibinfo  {journal} {Phys. Rev. Lett.}\ }\textbf {\bibinfo {volume} {99}},\
  \bibinfo {pages} {227602} (\bibinfo {year} {2007})}\BibitemShut {NoStop}%
\bibitem [{\citenamefont {Zhou}\ and\ \citenamefont
  {Kennedy}(2006{\natexlab{a}})}]{cmo_lattice}%
  \BibitemOpen
  \bibfield  {author} {\bibinfo {author} {\bibfnamefont {Q.}~\bibnamefont
  {Zhou}}\ and\ \bibinfo {author} {\bibfnamefont {B.~J.}\ \bibnamefont
  {Kennedy}},\ }\bibfield  {title} {\bibinfo {title} {Thermal expansion and
  structure of orthorhombic {CaMnO$_3$}},\ }\href
  {https://doi.org/https://doi.org/10.1016/j.jpcs.2006.02.011} {\bibfield
  {journal} {\bibinfo  {journal} {Journal of Physics and Chemistry of Solids}\
  }\textbf {\bibinfo {volume} {67}},\ \bibinfo {pages} {1595} (\bibinfo {year}
  {2006}{\natexlab{a}})}\BibitemShut {NoStop}%
\bibitem [{\citenamefont {Momma}\ and\ \citenamefont {Izumi}(2011)}]{VESTA}%
  \BibitemOpen
  \bibfield  {author} {\bibinfo {author} {\bibfnamefont {K.}~\bibnamefont
  {Momma}}\ and\ \bibinfo {author} {\bibfnamefont {F.}~\bibnamefont {Izumi}},\
  }\bibfield  {title} {\bibinfo {title} {{VESTA 3} for three-dimensional
  visualization of crystal, volumetric and morphology data},\ }\href
  {https://doi.org/10.1107/S0021889811038970} {\bibfield  {journal} {\bibinfo
  {journal} {J. Appl. Cryst.}\ }\textbf {\bibinfo {volume} {44}},\ \bibinfo
  {pages} {1272} (\bibinfo {year} {2011})}\BibitemShut {NoStop}%
\bibitem [{\citenamefont {Aroyo}\ \emph
  {et~al.}(2006{\natexlab{a}})\citenamefont {Aroyo}, \citenamefont
  {Perez-Mato}, \citenamefont {Capillas}, \citenamefont {Kroumova},
  \citenamefont {Ivantchev}, \citenamefont {Madariaga}, \citenamefont {Kirov},\
  and\ \citenamefont {Wondratschek}}]{Bilbao2006}%
  \BibitemOpen
  \bibfield  {author} {\bibinfo {author} {\bibfnamefont {M.~I.}\ \bibnamefont
  {Aroyo}}, \bibinfo {author} {\bibfnamefont {J.~M.}\ \bibnamefont
  {Perez-Mato}}, \bibinfo {author} {\bibfnamefont {C.}~\bibnamefont
  {Capillas}}, \bibinfo {author} {\bibfnamefont {E.}~\bibnamefont {Kroumova}},
  \bibinfo {author} {\bibfnamefont {S.}~\bibnamefont {Ivantchev}}, \bibinfo
  {author} {\bibfnamefont {G.}~\bibnamefont {Madariaga}}, \bibinfo {author}
  {\bibfnamefont {A.}~\bibnamefont {Kirov}},\ and\ \bibinfo {author}
  {\bibfnamefont {H.}~\bibnamefont {Wondratschek}},\ }\bibfield  {title}
  {\bibinfo {title} {{Bilbao Crystallographic Server: I. Databases and
  crystallographic computing programs}},\ }\href
  {https://doi.org/doi:10.1524/zkri.2006.221.1.15} {\bibfield  {journal}
  {\bibinfo  {journal} {Zeitschrift für Kristallographie - Crystalline
  Materials}\ }\textbf {\bibinfo {volume} {221}},\ \bibinfo {pages} {15}
  (\bibinfo {year} {2006}{\natexlab{a}})}\BibitemShut {NoStop}%
\bibitem [{\citenamefont {Zhou}\ and\ \citenamefont
  {Kennedy}(2006{\natexlab{b}})}]{cmo_wyckoff}%
  \BibitemOpen
  \bibfield  {author} {\bibinfo {author} {\bibfnamefont {Q.}~\bibnamefont
  {Zhou}}\ and\ \bibinfo {author} {\bibfnamefont {B.~J.}\ \bibnamefont
  {Kennedy}},\ }\bibfield  {title} {\bibinfo {title} {Thermal expansion and
  structure of orthorhombic {CaMnO$_3$}},\ }\href
  {https://doi.org/https://doi.org/10.1016/j.jpcs.2006.02.011} {\bibfield
  {journal} {\bibinfo  {journal} {Journal of Physics and Chemistry of Solids}\
  }\textbf {\bibinfo {volume} {67}},\ \bibinfo {pages} {1595} (\bibinfo {year}
  {2006}{\natexlab{b}})}\BibitemShut {NoStop}%
\bibitem [{\citenamefont {Kresse}\ and\ \citenamefont
  {Furthm{\"{u}}ller}(1996)}]{vasp1}%
  \BibitemOpen
  \bibfield  {author} {\bibinfo {author} {\bibfnamefont {G.}~\bibnamefont
  {Kresse}}\ and\ \bibinfo {author} {\bibfnamefont {J.}~\bibnamefont
  {Furthm{\"{u}}ller}},\ }\bibfield  {title} {\bibinfo {title} {{Efficient
  iterative schemes for ab initio total-energy calculations using a plane-wave
  basis set}},\ }\href {https://doi.org/10.1103/PhysRevB.54.11169} {\bibfield
  {journal} {\bibinfo  {journal} {Physical Review B}\ }\textbf {\bibinfo
  {volume} {54}},\ \bibinfo {pages} {11169} (\bibinfo {year}
  {1996})}\BibitemShut {NoStop}%
\bibitem [{\citenamefont {Kresse}\ and\ \citenamefont {Joubert}(1999)}]{vasp2}%
  \BibitemOpen
  \bibfield  {author} {\bibinfo {author} {\bibfnamefont {G.}~\bibnamefont
  {Kresse}}\ and\ \bibinfo {author} {\bibfnamefont {D.}~\bibnamefont
  {Joubert}},\ }\bibfield  {title} {\bibinfo {title} {{From ultrasoft
  pseudopotentials to the projector augmented-wave method}},\ }\href
  {https://doi.org/10.1103/PhysRevB.59.1758} {\bibfield  {journal} {\bibinfo
  {journal} {Physical Review B}\ }\textbf {\bibinfo {volume} {59}},\ \bibinfo
  {pages} {1758} (\bibinfo {year} {1999})}\BibitemShut {NoStop}%
\bibitem [{\citenamefont {Bl{\"{o}}chl}(1994)}]{paw}%
  \BibitemOpen
  \bibfield  {author} {\bibinfo {author} {\bibfnamefont {P.~E.}\ \bibnamefont
  {Bl{\"{o}}chl}},\ }\bibfield  {title} {\bibinfo {title} {{Projector
  augmented-wave method}},\ }\href {https://doi.org/10.1103/PhysRevB.50.17953}
  {\bibfield  {journal} {\bibinfo  {journal} {Phys. Rev. B}\ }\textbf {\bibinfo
  {volume} {50}},\ \bibinfo {pages} {17953} (\bibinfo {year}
  {1994})}\BibitemShut {NoStop}%
\bibitem [{\citenamefont {Perdew}\ \emph {et~al.}(1996)\citenamefont {Perdew},
  \citenamefont {Burke},\ and\ \citenamefont {Ernzerhof}}]{pbe}%
  \BibitemOpen
  \bibfield  {author} {\bibinfo {author} {\bibfnamefont {J.~P.}\ \bibnamefont
  {Perdew}}, \bibinfo {author} {\bibfnamefont {K.}~\bibnamefont {Burke}},\ and\
  \bibinfo {author} {\bibfnamefont {M.}~\bibnamefont {Ernzerhof}},\ }\bibfield
  {title} {\bibinfo {title} {{Generalized Gradient Approximation Made
  Simple}},\ }\href {https://doi.org/10.1103/PhysRevLett.77.3865} {\bibfield
  {journal} {\bibinfo  {journal} {Phys. Rev. Lett.}\ }\textbf {\bibinfo
  {volume} {77}},\ \bibinfo {pages} {3865} (\bibinfo {year}
  {1996})}\BibitemShut {NoStop}%
\bibitem [{\citenamefont {Dudarev}\ \emph {et~al.}(1998)\citenamefont
  {Dudarev}, \citenamefont {Botton}, \citenamefont {Savrasov}, \citenamefont
  {Humphreys},\ and\ \citenamefont {Sutton}}]{DudarevPRB98}%
  \BibitemOpen
  \bibfield  {author} {\bibinfo {author} {\bibfnamefont {S.~L.}\ \bibnamefont
  {Dudarev}}, \bibinfo {author} {\bibfnamefont {G.~A.}\ \bibnamefont {Botton}},
  \bibinfo {author} {\bibfnamefont {S.~Y.}\ \bibnamefont {Savrasov}}, \bibinfo
  {author} {\bibfnamefont {C.~J.}\ \bibnamefont {Humphreys}},\ and\ \bibinfo
  {author} {\bibfnamefont {A.~P.}\ \bibnamefont {Sutton}},\ }\bibfield  {title}
  {\bibinfo {title} {{Electron-energy-loss spectra and the structural stability
  of nickel oxide: An LSDA+U study}},\ }\href
  {https://doi.org/10.1103/PhysRevB.57.1505} {\bibfield  {journal} {\bibinfo
  {journal} {Phys. Rev. B}\ }\textbf {\bibinfo {volume} {57}},\ \bibinfo
  {pages} {1505} (\bibinfo {year} {1998})}\BibitemShut {NoStop}%
\bibitem [{\citenamefont {Bae}\ \emph {et~al.}(2017)\citenamefont {Bae},
  \citenamefont {Kov{\'{a}}cs}, \citenamefont {Zhao}, \citenamefont
  {{\'{I}}{\~{n}}iguez}, \citenamefont {Yasui}, \citenamefont {Ichinose},\ and\
  \citenamefont {Naganuma}}]{BFO_ueff}%
  \BibitemOpen
  \bibfield  {author} {\bibinfo {author} {\bibfnamefont {I.~T.}\ \bibnamefont
  {Bae}}, \bibinfo {author} {\bibfnamefont {A.}~\bibnamefont {Kov{\'{a}}cs}},
  \bibinfo {author} {\bibfnamefont {H.~J.}\ \bibnamefont {Zhao}}, \bibinfo
  {author} {\bibfnamefont {J.}~\bibnamefont {{\'{I}}{\~{n}}iguez}}, \bibinfo
  {author} {\bibfnamefont {S.}~\bibnamefont {Yasui}}, \bibinfo {author}
  {\bibfnamefont {T.}~\bibnamefont {Ichinose}},\ and\ \bibinfo {author}
  {\bibfnamefont {H.}~\bibnamefont {Naganuma}},\ }\bibfield  {title} {\bibinfo
  {title} {Elucidation of crystal and electronic structures within highly
  strained {BiFeO$_3$} by transmission electron microscopy and first-principles
  simulation},\ }\href {https://doi.org/10.1038/srep46498} {\bibfield
  {journal} {\bibinfo  {journal} {Scientific Reports}\ }\textbf {\bibinfo
  {volume} {7}},\ \bibinfo {pages} {46498} (\bibinfo {year}
  {2017})}\BibitemShut {NoStop}%
\bibitem [{\citenamefont {Aschauer}\ \emph {et~al.}(2013)\citenamefont
  {Aschauer}, \citenamefont {Pfenninger}, \citenamefont {Selbach},
  \citenamefont {Grande},\ and\ \citenamefont {Spaldin}}]{CMO_ueff}%
  \BibitemOpen
  \bibfield  {author} {\bibinfo {author} {\bibfnamefont {U.}~\bibnamefont
  {Aschauer}}, \bibinfo {author} {\bibfnamefont {R.}~\bibnamefont
  {Pfenninger}}, \bibinfo {author} {\bibfnamefont {S.~M.}\ \bibnamefont
  {Selbach}}, \bibinfo {author} {\bibfnamefont {T.}~\bibnamefont {Grande}},\
  and\ \bibinfo {author} {\bibfnamefont {N.~A.}\ \bibnamefont {Spaldin}},\
  }\bibfield  {title} {\bibinfo {title} {Strain-controlled oxygen vacancy
  formation and ordering in {CaMnO$_3$}},\ }\href
  {https://doi.org/10.1103/PhysRevB.88.054111} {\bibfield  {journal} {\bibinfo
  {journal} {Phys. Rev. B}\ }\textbf {\bibinfo {volume} {88}},\ \bibinfo
  {pages} {054111} (\bibinfo {year} {2013})}\BibitemShut {NoStop}%
\bibitem [{\citenamefont {Bl\"ochl}\ \emph {et~al.}(1994)\citenamefont
  {Bl\"ochl}, \citenamefont {Jepsen},\ and\ \citenamefont
  {Andersen}}]{BlochTetrahedron94}%
  \BibitemOpen
  \bibfield  {author} {\bibinfo {author} {\bibfnamefont {P.~E.}\ \bibnamefont
  {Bl\"ochl}}, \bibinfo {author} {\bibfnamefont {O.}~\bibnamefont {Jepsen}},\
  and\ \bibinfo {author} {\bibfnamefont {O.~K.}\ \bibnamefont {Andersen}},\
  }\bibfield  {title} {\bibinfo {title} {{Improved tetrahedron method for
  Brillouin-zone integrations}},\ }\href
  {https://doi.org/10.1103/PhysRevB.49.16223} {\bibfield  {journal} {\bibinfo
  {journal} {Phys. Rev. B}\ }\textbf {\bibinfo {volume} {49}},\ \bibinfo
  {pages} {16223} (\bibinfo {year} {1994})}\BibitemShut {NoStop}%
\bibitem [{\citenamefont {Pizzi}\ \emph {et~al.}(2020)\citenamefont {Pizzi},
  \citenamefont {Vitale}, \citenamefont {Arita}, \citenamefont {Bl{\"{u}}gel},
  \citenamefont {Freimuth}, \citenamefont {G{\'{e}}ranton}, \citenamefont
  {Gibertini}, \citenamefont {Gresch}, \citenamefont {Johnson}, \citenamefont
  {Koretsune}, \citenamefont {Iba{\~{n}}ez-Azpiroz}, \citenamefont {Lee},
  \citenamefont {Lihm}, \citenamefont {Marchand}, \citenamefont {Marrazzo},
  \citenamefont {Mokrousov}, \citenamefont {Mustafa}, \citenamefont {Nohara},
  \citenamefont {Nomura}, \citenamefont {Paulatto}, \citenamefont
  {Ponc{\'{e}}}, \citenamefont {Ponweiser}, \citenamefont {Qiao}, \citenamefont
  {Th{\"{o}}le}, \citenamefont {Tsirkin}, \citenamefont {Wierzbowska},
  \citenamefont {Marzari}, \citenamefont {Vanderbilt}, \citenamefont {Souza},
  \citenamefont {Mostofi},\ and\ \citenamefont {Yates}}]{wannier90}%
  \BibitemOpen
  \bibfield  {author} {\bibinfo {author} {\bibfnamefont {G.}~\bibnamefont
  {Pizzi}}, \bibinfo {author} {\bibfnamefont {V.}~\bibnamefont {Vitale}},
  \bibinfo {author} {\bibfnamefont {R.}~\bibnamefont {Arita}}, \bibinfo
  {author} {\bibfnamefont {S.}~\bibnamefont {Bl{\"{u}}gel}}, \bibinfo {author}
  {\bibfnamefont {F.}~\bibnamefont {Freimuth}}, \bibinfo {author}
  {\bibfnamefont {G.}~\bibnamefont {G{\'{e}}ranton}}, \bibinfo {author}
  {\bibfnamefont {M.}~\bibnamefont {Gibertini}}, \bibinfo {author}
  {\bibfnamefont {D.}~\bibnamefont {Gresch}}, \bibinfo {author} {\bibfnamefont
  {C.}~\bibnamefont {Johnson}}, \bibinfo {author} {\bibfnamefont
  {T.}~\bibnamefont {Koretsune}}, \bibinfo {author} {\bibfnamefont
  {J.}~\bibnamefont {Iba{\~{n}}ez-Azpiroz}}, \bibinfo {author} {\bibfnamefont
  {H.}~\bibnamefont {Lee}}, \bibinfo {author} {\bibfnamefont {J.-M.}\
  \bibnamefont {Lihm}}, \bibinfo {author} {\bibfnamefont {D.}~\bibnamefont
  {Marchand}}, \bibinfo {author} {\bibfnamefont {A.}~\bibnamefont {Marrazzo}},
  \bibinfo {author} {\bibfnamefont {Y.}~\bibnamefont {Mokrousov}}, \bibinfo
  {author} {\bibfnamefont {J.~I.}\ \bibnamefont {Mustafa}}, \bibinfo {author}
  {\bibfnamefont {Y.}~\bibnamefont {Nohara}}, \bibinfo {author} {\bibfnamefont
  {Y.}~\bibnamefont {Nomura}}, \bibinfo {author} {\bibfnamefont
  {L.}~\bibnamefont {Paulatto}}, \bibinfo {author} {\bibfnamefont
  {S.}~\bibnamefont {Ponc{\'{e}}}}, \bibinfo {author} {\bibfnamefont
  {T.}~\bibnamefont {Ponweiser}}, \bibinfo {author} {\bibfnamefont
  {J.}~\bibnamefont {Qiao}}, \bibinfo {author} {\bibfnamefont {F.}~\bibnamefont
  {Th{\"{o}}le}}, \bibinfo {author} {\bibfnamefont {S.~S.}\ \bibnamefont
  {Tsirkin}}, \bibinfo {author} {\bibfnamefont {M.}~\bibnamefont
  {Wierzbowska}}, \bibinfo {author} {\bibfnamefont {N.}~\bibnamefont
  {Marzari}}, \bibinfo {author} {\bibfnamefont {D.}~\bibnamefont {Vanderbilt}},
  \bibinfo {author} {\bibfnamefont {I.}~\bibnamefont {Souza}}, \bibinfo
  {author} {\bibfnamefont {A.~A.}\ \bibnamefont {Mostofi}},\ and\ \bibinfo
  {author} {\bibfnamefont {J.~R.}\ \bibnamefont {Yates}},\ }\bibfield  {title}
  {\bibinfo {title} {{Wannier90 as a community code: new features and
  applications}},\ }\href {https://doi.org/10.1088/1361-648X/ab51ff} {\bibfield
   {journal} {\bibinfo  {journal} {J. Phys.: Condens. Matter}\ }\textbf
  {\bibinfo {volume} {32}},\ \bibinfo {pages} {165902} (\bibinfo {year}
  {2020})}\BibitemShut {NoStop}%
\bibitem [{\citenamefont {Bharamagoudar}\ \emph {et~al.}(2022)\citenamefont
  {Bharamagoudar}, \citenamefont {{Angadi V}}, \citenamefont {Pattar},
  \citenamefont {Patil}, \citenamefont {Patil}, \citenamefont {S},
  \citenamefont {Kulkarni}, \citenamefont {Malakannavar},\ and\ \citenamefont
  {Matteppanavar}}]{CMO_ordering_temp}%
  \BibitemOpen
  \bibfield  {author} {\bibinfo {author} {\bibfnamefont {R.}~\bibnamefont
  {Bharamagoudar}}, \bibinfo {author} {\bibfnamefont {J.}~\bibnamefont {{Angadi
  V}}}, \bibinfo {author} {\bibfnamefont {V.}~\bibnamefont {Pattar}}, \bibinfo
  {author} {\bibfnamefont {A.~S.}\ \bibnamefont {Patil}}, \bibinfo {author}
  {\bibfnamefont {S.}~\bibnamefont {Patil}}, \bibinfo {author} {\bibfnamefont
  {R.}~\bibnamefont {S}}, \bibinfo {author} {\bibfnamefont {S.}~\bibnamefont
  {Kulkarni}}, \bibinfo {author} {\bibfnamefont {M.~V.}\ \bibnamefont
  {Malakannavar}},\ and\ \bibinfo {author} {\bibfnamefont {S.}~\bibnamefont
  {Matteppanavar}},\ }\bibfield  {title} {\bibinfo {title} {Low temperature
  magnetic properties of {Gd} doped {CaMnO$_3$}},\ }\href
  {https://doi.org/https://doi.org/10.1016/j.cdc.2022.100846} {\bibfield
  {journal} {\bibinfo  {journal} {Chemical Data Collections}\ }\textbf
  {\bibinfo {volume} {39}},\ \bibinfo {pages} {100846} (\bibinfo {year}
  {2022})}\BibitemShut {NoStop}%
\bibitem [{\citenamefont {Arnold}\ \emph {et~al.}(2009)\citenamefont {Arnold},
  \citenamefont {Knight}, \citenamefont {Morrison},\ and\ \citenamefont
  {Lightfoot}}]{BFOFerroPara}%
  \BibitemOpen
  \bibfield  {author} {\bibinfo {author} {\bibfnamefont {D.~C.}\ \bibnamefont
  {Arnold}}, \bibinfo {author} {\bibfnamefont {K.~S.}\ \bibnamefont {Knight}},
  \bibinfo {author} {\bibfnamefont {F.~D.}\ \bibnamefont {Morrison}},\ and\
  \bibinfo {author} {\bibfnamefont {P.}~\bibnamefont {Lightfoot}},\ }\bibfield
  {title} {\bibinfo {title} {Ferroelectric-paraelectric transition in
  {BiFeO$_3$}: Crystal structure of the orthorhombic $\beta$ phase},\ }\href
  {https://doi.org/10.1103/PhysRevLett.102.027602} {\bibfield  {journal}
  {\bibinfo  {journal} {Phys. Rev. Lett.}\ }\textbf {\bibinfo {volume} {102}},\
  \bibinfo {pages} {027602} (\bibinfo {year} {2009})}\BibitemShut {NoStop}%
\bibitem [{\citenamefont {Mužević}\ \emph {et~al.}(2023)\citenamefont
  {Mužević}, \citenamefont {Lukačević}, \citenamefont {Kovač},
  \citenamefont {Gracin}, \citenamefont {Žužić}, \citenamefont {Macan},\
  and\ \citenamefont {Pajtler}}]{CMO_bandgap2}%
  \BibitemOpen
  \bibfield  {author} {\bibinfo {author} {\bibfnamefont {M.}~\bibnamefont
  {Mužević}}, \bibinfo {author} {\bibfnamefont {I.}~\bibnamefont
  {Lukačević}}, \bibinfo {author} {\bibfnamefont {I.}~\bibnamefont {Kovač}},
  \bibinfo {author} {\bibfnamefont {D.}~\bibnamefont {Gracin}}, \bibinfo
  {author} {\bibfnamefont {A.}~\bibnamefont {Žužić}}, \bibinfo {author}
  {\bibfnamefont {J.}~\bibnamefont {Macan}},\ and\ \bibinfo {author}
  {\bibfnamefont {M.~V.}\ \bibnamefont {Pajtler}},\ }\bibfield  {title}
  {\bibinfo {title} {Potential of {AMnO$_3$} {(A=Ca, Sr, Ba, La)} as active
  layer in inorganic perovskite solar cells},\ }\href
  {https://doi.org/https://doi.org/10.1002/cphc.202200837} {\bibfield
  {journal} {\bibinfo  {journal} {ChemPhysChem}\ }\textbf {\bibinfo {volume}
  {24}},\ \bibinfo {pages} {e202200837} (\bibinfo {year} {2023})}\BibitemShut
  {NoStop}%
\bibitem [{\citenamefont {Molinari}\ \emph {et~al.}(2014)\citenamefont
  {Molinari}, \citenamefont {Tompsett}, \citenamefont {Parker}, \citenamefont
  {Azough},\ and\ \citenamefont {Freer}}]{CMO_magmom}%
  \BibitemOpen
  \bibfield  {author} {\bibinfo {author} {\bibfnamefont {M.}~\bibnamefont
  {Molinari}}, \bibinfo {author} {\bibfnamefont {D.~A.}\ \bibnamefont
  {Tompsett}}, \bibinfo {author} {\bibfnamefont {S.~C.}\ \bibnamefont
  {Parker}}, \bibinfo {author} {\bibfnamefont {F.}~\bibnamefont {Azough}},\
  and\ \bibinfo {author} {\bibfnamefont {R.}~\bibnamefont {Freer}},\ }\bibfield
   {title} {\bibinfo {title} {Structural, electronic and thermoelectric
  behaviour of {CaMnO$_3$}and {CaMnO$_{(3 - \delta)}$}},\ }\href
  {https://doi.org/10.1039/C4TA01514B} {\bibfield  {journal} {\bibinfo
  {journal} {J. Mater. Chem. A}\ }\textbf {\bibinfo {volume} {2}},\ \bibinfo
  {pages} {14109} (\bibinfo {year} {2014})}\BibitemShut {NoStop}%
\bibitem [{\citenamefont {Fernandes}\ \emph {et~al.}(2024)\citenamefont
  {Fernandes}, \citenamefont {{De Carvalho}}, \citenamefont {Birol},\ and\
  \citenamefont {Pereira}}]{Fernandes2024}%
  \BibitemOpen
  \bibfield  {author} {\bibinfo {author} {\bibfnamefont {R.~M.}\ \bibnamefont
  {Fernandes}}, \bibinfo {author} {\bibfnamefont {V.~S.}\ \bibnamefont {{De
  Carvalho}}}, \bibinfo {author} {\bibfnamefont {T.}~\bibnamefont {Birol}},\
  and\ \bibinfo {author} {\bibfnamefont {R.~G.}\ \bibnamefont {Pereira}},\
  }\bibfield  {title} {\bibinfo {title} {{Topological transition from nodal to
  nodeless Zeeman splitting in altermagnets}},\ }\href
  {https://doi.org/10.1103/PhysRevB.109.024404} {\bibfield  {journal} {\bibinfo
   {journal} {Physical Review B}\ }\textbf {\bibinfo {volume} {109}},\ \bibinfo
  {pages} {024404} (\bibinfo {year} {2024})}\BibitemShut {NoStop}%
\bibitem [{\citenamefont {Aroyo}\ \emph
  {et~al.}(2006{\natexlab{b}})\citenamefont {Aroyo}, \citenamefont {Kirov},
  \citenamefont {Capillas}, \citenamefont {Perez-Mato},\ and\ \citenamefont
  {Wondratschek}}]{Aroyo:xo5013}%
  \BibitemOpen
  \bibfield  {author} {\bibinfo {author} {\bibfnamefont {M.~I.}\ \bibnamefont
  {Aroyo}}, \bibinfo {author} {\bibfnamefont {A.}~\bibnamefont {Kirov}},
  \bibinfo {author} {\bibfnamefont {C.}~\bibnamefont {Capillas}}, \bibinfo
  {author} {\bibfnamefont {J.~M.}\ \bibnamefont {Perez-Mato}},\ and\ \bibinfo
  {author} {\bibfnamefont {H.}~\bibnamefont {Wondratschek}},\ }\bibfield
  {title} {\bibinfo {title} {{Bilbao Crystallographic Server. II.
  Representations of crystallographic point groups and space groups}},\ }\href
  {https://doi.org/10.1107/S0108767305040286} {\bibfield  {journal} {\bibinfo
  {journal} {Acta Crystallographica Section A}\ }\textbf {\bibinfo {volume}
  {62}},\ \bibinfo {pages} {115} (\bibinfo {year}
  {2006}{\natexlab{b}})}\BibitemShut {NoStop}%
\bibitem [{\citenamefont {Dresselhaus}\ \emph {et~al.}(2008)\citenamefont
  {Dresselhaus}, \citenamefont {Dresselhaus},\ and\ \citenamefont
  {Jorio}}]{DresselhausGroupTheory08}%
  \BibitemOpen
  \bibfield  {author} {\bibinfo {author} {\bibfnamefont {M.~S.}\ \bibnamefont
  {Dresselhaus}}, \bibinfo {author} {\bibfnamefont {G.}~\bibnamefont
  {Dresselhaus}},\ and\ \bibinfo {author} {\bibfnamefont {A.}~\bibnamefont
  {Jorio}},\ }\href {https://doi.org/10.1007/978-3-540-32899-5} {\emph
  {\bibinfo {title} {{Group Theory: Application to the Physics of Condensed
  Matter}}}}\ (\bibinfo  {publisher} {Springer Berlin Heidelberg},\ \bibinfo
  {address} {Berlin, Heidelberg},\ \bibinfo {year} {2008})\BibitemShut
  {NoStop}%
\bibitem [{\citenamefont {Perez-Mato}\ \emph
  {et~al.}(2015{\natexlab{a}})\citenamefont {Perez-Mato}, \citenamefont
  {Gallego}, \citenamefont {Tasci}, \citenamefont {Elcoro}, \citenamefont
  {de~la Flor},\ and\ \citenamefont {Aroyo}}]{MSGbilbao2015}%
  \BibitemOpen
  \bibfield  {author} {\bibinfo {author} {\bibfnamefont {J.}~\bibnamefont
  {Perez-Mato}}, \bibinfo {author} {\bibfnamefont {S.}~\bibnamefont {Gallego}},
  \bibinfo {author} {\bibfnamefont {E.}~\bibnamefont {Tasci}}, \bibinfo
  {author} {\bibfnamefont {L.}~\bibnamefont {Elcoro}}, \bibinfo {author}
  {\bibfnamefont {G.}~\bibnamefont {de~la Flor}},\ and\ \bibinfo {author}
  {\bibfnamefont {M.}~\bibnamefont {Aroyo}},\ }\bibfield  {title} {\bibinfo
  {title} {Symmetry-based computational tools for magnetic crystallography},\
  }\href {https://doi.org/https://doi.org/10.1146/annurev-matsci-070214-021008}
  {\bibfield  {journal} {\bibinfo  {journal} {Annual Review of Materials
  Research}\ }\textbf {\bibinfo {volume} {45}},\ \bibinfo {pages} {217}
  (\bibinfo {year} {2015}{\natexlab{a}})}\BibitemShut {NoStop}%
\bibitem [{\citenamefont {Bradley}\ and\ \citenamefont
  {Cracknell}(2009)}]{BradleySymmetrySolid}%
  \BibitemOpen
  \bibfield  {author} {\bibinfo {author} {\bibfnamefont {C.~J.}\ \bibnamefont
  {Bradley}}\ and\ \bibinfo {author} {\bibfnamefont {A.~P.}\ \bibnamefont
  {Cracknell}},\ }\href@noop {} {\emph {\bibinfo {title} {{The Mathematical
  Theory of Symmetry in Solids: Representation Theory for Point Groups and
  Space Groups}}}}\ (\bibinfo  {publisher} {Oxford University Press Inc.},\
  \bibinfo {year} {2009})\BibitemShut {NoStop}%
\bibitem [{\citenamefont {Perez-Mato}\ \emph
  {et~al.}(2015{\natexlab{b}})\citenamefont {Perez-Mato}, \citenamefont
  {Gallego}, \citenamefont {Tasci}, \citenamefont {Elcoro}, \citenamefont
  {de~la Flor},\ and\ \citenamefont {Aroyo}}]{MagneticSpaceGroup}%
  \BibitemOpen
  \bibfield  {author} {\bibinfo {author} {\bibfnamefont {J.}~\bibnamefont
  {Perez-Mato}}, \bibinfo {author} {\bibfnamefont {S.}~\bibnamefont {Gallego}},
  \bibinfo {author} {\bibfnamefont {E.}~\bibnamefont {Tasci}}, \bibinfo
  {author} {\bibfnamefont {L.}~\bibnamefont {Elcoro}}, \bibinfo {author}
  {\bibfnamefont {G.}~\bibnamefont {de~la Flor}},\ and\ \bibinfo {author}
  {\bibfnamefont {M.}~\bibnamefont {Aroyo}},\ }\bibfield  {title} {\bibinfo
  {title} {Symmetry-based computational tools for magnetic crystallography},\
  }\href {https://doi.org/https://doi.org/10.1146/annurev-matsci-070214-021008}
  {\bibfield  {journal} {\bibinfo  {journal} {Annual Review of Materials
  Research}\ }\textbf {\bibinfo {volume} {45}},\ \bibinfo {pages} {217}
  (\bibinfo {year} {2015}{\natexlab{b}})}\BibitemShut {NoStop}%
\bibitem [{\citenamefont {{\v{S}}mejkal}\ \emph {et~al.}(2022)\citenamefont
  {{\v{S}}mejkal}, \citenamefont {Sinova},\ and\ \citenamefont
  {Jungwirth}}]{SmejkalPRX22}%
  \BibitemOpen
  \bibfield  {author} {\bibinfo {author} {\bibfnamefont {L.}~\bibnamefont
  {{\v{S}}mejkal}}, \bibinfo {author} {\bibfnamefont {J.}~\bibnamefont
  {Sinova}},\ and\ \bibinfo {author} {\bibfnamefont {T.}~\bibnamefont
  {Jungwirth}},\ }\bibfield  {title} {\bibinfo {title} {{Beyond Conventional
  Ferromagnetism and Antiferromagnetism: A Phase with Nonrelativistic Spin and
  Crystal Rotation Symmetry}},\ }\href
  {https://doi.org/10.1103/PhysRevX.12.031042} {\bibfield  {journal} {\bibinfo
  {journal} {Phys. Rev. X}\ }\textbf {\bibinfo {volume} {12}},\ \bibinfo
  {pages} {031042} (\bibinfo {year} {2022})}\BibitemShut {NoStop}%
\bibitem [{\citenamefont {Smejkal}\ \emph {et~al.}()\citenamefont {Smejkal},
  \citenamefont {MacDonald}, \citenamefont {Sinova}, \citenamefont
  {Nakatsuji},\ and\ \citenamefont {Jungwirth}}]{Smejkal2021AHE}%
  \BibitemOpen
  \bibfield  {author} {\bibinfo {author} {\bibfnamefont {L.}~\bibnamefont
  {Smejkal}}, \bibinfo {author} {\bibfnamefont {A.~H.}\ \bibnamefont
  {MacDonald}}, \bibinfo {author} {\bibfnamefont {J.}~\bibnamefont {Sinova}},
  \bibinfo {author} {\bibfnamefont {S.}~\bibnamefont {Nakatsuji}},\ and\
  \bibinfo {author} {\bibfnamefont {T.}~\bibnamefont {Jungwirth}},\ }\bibfield
  {title} {\bibinfo {title} {Anomalous hall antiferromagnets},\ }\href
  {https://arxiv.org/abs/2107.03321} {\ }\Eprint
  {https://arxiv.org/abs/2107.03321 (2021)} {arXiv:2107.03321 (2021)}
  \BibitemShut {NoStop}%
\end{thebibliography}


\end{document}